\documentclass[a4paper,11pt]{article}
\pdfoutput=1
\usepackage{jinstpub}
\usepackage{indentfirst}
\usepackage{amsmath}
\usepackage{tikz}
\usetikzlibrary{positioning}
\usetikzlibrary{arrows.meta}
\usepackage{adjustbox}
\usepackage{multirow}
\usepackage{subfigure}
\usepackage{makecell}
\usepackage{siunitx}

\newcommand{\Ar}{\mathrm{Ar}}
\newcommand{\Xe}{\mathrm{Xe}}

\title{Pulse shape study of the fast scintillation light emitted from xenon-doped liquid argon using silicon photomultipliers}
\author[a]{C.~Galbiati,}
\author[a,1]{X.~Li,\note{Corresponding authors.}}
\author[a,2]{J.~Luo,\note{Present address: Department of Physics, Brown University, Providence, Rhode Island 02912, USA.}}
\author[a]{D.~R.~Marlow}
\author[b]{H.~Wang}
\author[c,1]{and Y.~Wang}

\affiliation[a]{Department of Physics, Princeton University\\New Jersey 08544, USA}
\affiliation[b]{Department of Physics and Astronomy, University of California, Los Angeles\\California 90095, USA}
\affiliation[c]{Institute of High Energy Physics, Chinese Academy of Sciences\\Beijing 100049, China}

\emailAdd{wangyi90@ihep.ac.cn}

\abstract{
Xenon-doped liquid argon has been proposed as a good alternative to pure liquid argon in scintillation detectors.   
In this study, we report on the measurement of the time profile of scintillation light emitted from xenon-doped liquid argon with molar concentrations up to \SI{1600}{ppm}.
A compact setup has been developed for this study, with silicon photomultiplier (SiPM) as the photosensor and $^{210}\mathrm{Po}$ and $^{90}\mathrm{Sr}$ as scintillation sources. 
An effective model based on the de-excitation processes has been developed to describe the data. 
The results show that xenon-doped liquid argon is a good fast scintillator and can be used in lieu of tetraphenyl butadiene (TPB) in a way that preserves its capability for particle identification via pulse shape discrimination (PSD).
}


\begin{document}

\maketitle
\flushbottom

\section{Introduction and motivation}

Noble liquids, especially liquid argon and liquid xenon, are widely used as scintillators in high energy physics experiments, such as direct dark matter searches~\cite{darkside,xenon,deap,xmass}, observations of neutrinoless double beta decay~\cite{exo200} and measurements of neutrino oscillations~\cite{DUNE}.

The wavelengths of scintillation light emitted from liquid xenon and liquid argon are \SI{178}{nm} and \SI{128}{nm}, respectively. 
Atoms excited by charged particles usually relax and form excimers before emitting scintillation light.
Two components with different decay constants can be found in the scintillation light: a fast component from singlet state excimers and a slow component from triplet state excimers. 
The decay constants of these two components are \SI{4.3}{ns} and \SI{22}{ns} for liquid xenon, \SI{7}{ns} and \SI{1.6}{\us} for liquid argon, respectively~\cite{hitachi1983IonDensityPSD}.
The shorter emission time associated with liquid xenon makes it a more suitable candidate for high-speed applications and fast-timing applications, for example, the fast electromagnetic calorimeter~\cite{MEG}.

The \SI{178}{nm} wavelength light can be directly detected by ordinary photosensors, but the \SI{128}{nm} light lies beyond their range.  For this reason, wavelength shifter (WLS) is widely used in liquid argon detectors to shift the wavelength of the scintillation light from \SI{128}{nm} to the visible region. 
The most common-used WLS is tetraphenyl butadiene (TPB), which shifts the wavelength to \SI{420}{nm}~\cite{TPB}.
Even though the WLS can solve the light detection problem of many liquid argon detectors, the deposition of the WLS material complicates the construction of the detector.
In addition, the fluorescence process of TPB has a decay time around \SI{1}{ns}, which delays the prompt photons by a similar time scale. 
Although it is fast enough for most light-detection experiments, the TPB fluorescence process can limit performance in detectors with sub-nanosecond timing resolution requirements.

On the other hand, liquid argon exhibits advantages over liquid xenon. 
The boiling point of argon (\SI{87}{K}) is lower than that of xenon (\SI{165}{K}), which makes liquid argon a good medium for silicon photomultipliers (SiPMs), whose dark count rate has an exponential dependence on temperature.  
In addition,  argon is significantly less expensive than xenon and its low boiling point simplifies the purification process.  
Liquid argon also has outstanding particle identification capability with the scintillation light pulse shape discrimination (PSD) technique. 
Nuclear recoil events and electron recoil events can be distinguished with high efficiency and high acceptance~\cite{Mckinsey2008ArPSD}. 
Due to the similar decay time of the singlet and triplet states, PSD in liquid xenon is less powerful~\cite{LUX2018XePSD}.

In order to combine the advantages of both substances, the idea of xenon-doped liquid argon has been considered.
This mixture emits scintillation light at a wavelength of \SI{178}{nm}  and operates at temperatures close to the argon boiling point.    Its triplet state decay time is shorter than the one of pure liquid argon.
With these benefits, xenon-doped liquid argon has the potential to be used in many applications, such as massive detectors used to search for nucleon decay, supernova neutrino bursts, and the direct detection of low mass dark matter.  Moreover, since the de-excitation process in the mixture can be accomplished with direct energy transfer from argon excimers to xenon and direct emission of xenon light, it will be faster than the fluorescence processes of WLS. 
Yet to be experimentally demonstrated, xenon-doped liquid argon also has the potential to achieve sub-nanosecond timing resolution.
It might be suitable for building a full body three-dimensional time-of-flight positron-electron tomography (3D TOF-PET) scanner, associated with a large surface coverage of SiPMs.

In this work, a compact system set up to study the scintillation light from xenon-doped liquid argon using SiPMs as the photosensors is described.  Measurements are carried out to understand the evolution of the scintillation light time profile with different doping concentrations.

\section{Scintillation process of xenon-doped liquid argon}

\begin{figure}
    \centering
\begin{tikzpicture}[
node/.style={rectangle, draw=red!60, fill=red!5, very thick, minimum size=7mm},
dashnode/.style={rectangle, draw=red!60, fill=red!5, very thick, dashed, minimum size=7mm},
dashLine/.style={-{Latex[length=3mm, width=1.5mm]}, dashed},
solidLine/.style={-{Latex[length=3.5mm, width=2mm]}, line width=1.5pt},
Line/.style={-{Latex[length=3mm, width=1.5mm]}},
]
\node[node]     (ArIon)                             {Ar$^{+}$};
\node[node]     (Ar2Ion)    [right = of ArIon]      {Ar$_{2}^{+}$};
\node[node]     (Ar3P1)     [below = 1.5cm of ArIon]      {Ar$^{*3}$P$_{1}$};
\node[node]     (Ar3P2)     [right = of Ar3P1]      {Ar$^{*3}$P$_{2}$};
\node[node]     (Ar2Prime3) [below = 1.5cm of Ar3P2]  {Ar$_{2}^{*3}\Sigma$};
\node[node]     (Ar2Prime1) [below = 1.5cm of Ar3P1]     {Ar$_{2}^{*1}\Sigma$};
\node[node]     (XeIon)     [right = of Ar2Prime3] {Xe$^{+}$};
\node[node]     (Xe2Ion)    [right = of XeIon]      {Xe$_{2}^{+}$};
\node[dashnode]     (ArXeIon)   [right = 1.5cm of Xe2Ion]     {ArXe$^{+}$};
\node[node]     (Xe3P1)   [below = 1.5cm of XeIon]  {Xe$^{*3}$P$_{1}$};
\node[node]     (Xe3P2)   [right = of Xe3P1]  {Xe$^{*3}$P$_{2}$};
\node[node]     (ArXePrime) [below = 1.5cm of ArXeIon]    {ArXe$^{*}$};
\node[node]     (Xe2Prime1) [below = 1.5cm of Xe3P1]    {Xe$_{2}^{*1}\Sigma$};
\node[node]     (Xe2Prime3) [below = 1.5cm of Xe3P2]  {Xe$_{2}^{*3}\Sigma$};
\node[node]     (2Xehv)     [below = of Xe2Prime1]  {2Xe+h$\nu$};
\node[node]     (2Arhv)     [left = 2cm of 2Xehv]  {2Ar+h$\nu$};
\node[dashnode]     (ArXehv)    [right = 2.5cm of 2Xehv]  {Ar+Xe+h$\nu$};
 
\draw[Line] (ArIon.south)     -- (Ar3P1.north); 
\draw[Line] (ArIon.east)      -- (Ar2Ion.west);
\draw[Line] (Ar2Ion.south)    -- (Ar3P1.north);
\draw[Line] (Ar3P1.south)   -- (Ar2Prime1.north);
\draw[Line] (Ar3P1.east)   -- (Ar3P2.west);
\draw[Line] (Ar3P2.south)   -- (Ar2Prime3.north);
\draw[Line] (Ar3P1.south) -- (XeIon.west);
\draw[Line] (Ar3P2.south) -- (XeIon.west);
\draw[Line] (Ar2Prime1.east) to[out=45,in=135] (XeIon.west);
\draw[Line] (Ar2Prime3.east) -- (XeIon.west);
\draw[solidLine] (XeIon.south)   -- (Xe3P1.north);
\draw[Line] (XeIon.east)   -- (Xe2Ion.west);
\draw[dashLine] (XeIon.east) to[out=45,in=135] (ArXeIon.west);
\draw[Line] (Ar2Prime1.south) -- (2Arhv.north);
\draw[Line] (Ar2Prime3.south)  -- (2Arhv.north);
\draw[Line] (Ar2Prime1.south) -- (Xe3P1.north);
\draw[Line] (Ar2Prime3.south) -- (Xe3P1.north);
\draw[Line] (Xe2Ion.south)   -- (Xe3P1.north);
\draw[dashLine] (ArXeIon.south)   -- (ArXePrime.north);
\draw[Line] (Xe3P1.east)to[out=45,in=135] (ArXePrime.west);
\draw[Line] (Xe3P2.east) -- (ArXePrime.west);
\draw[solidLine] (Xe3P1.south) -- (Xe2Prime1.north);
\draw[Line] (Xe3P1.east) -- (Xe3P2.west);
\draw[Line] (ArXePrime.south)to[out=-90,in=20] (Xe2Prime1.north);
\draw[Line] (ArXePrime.south)to[out=-90,in=20] (Xe2Prime3.north);
\draw[Line] (Xe3P2.south)-- (Xe2Prime3.north);
\draw[dashLine] (ArXePrime.south)-- (ArXehv.north);
\draw[solidLine] (Xe2Prime1.south) -- (2Xehv.north);
\draw[Line] (Xe2Prime3.south) -- (2Xehv.north);
\end{tikzpicture}
\caption{Block diagram of the energy transfer in xenon-doped liquid argon. The solid arrows indicate the fastest pathway of energy release in scintillation light. The dashed arrows indicate the processes that have not yet been demonstrated.}
\label{fig:process}
\end{figure}
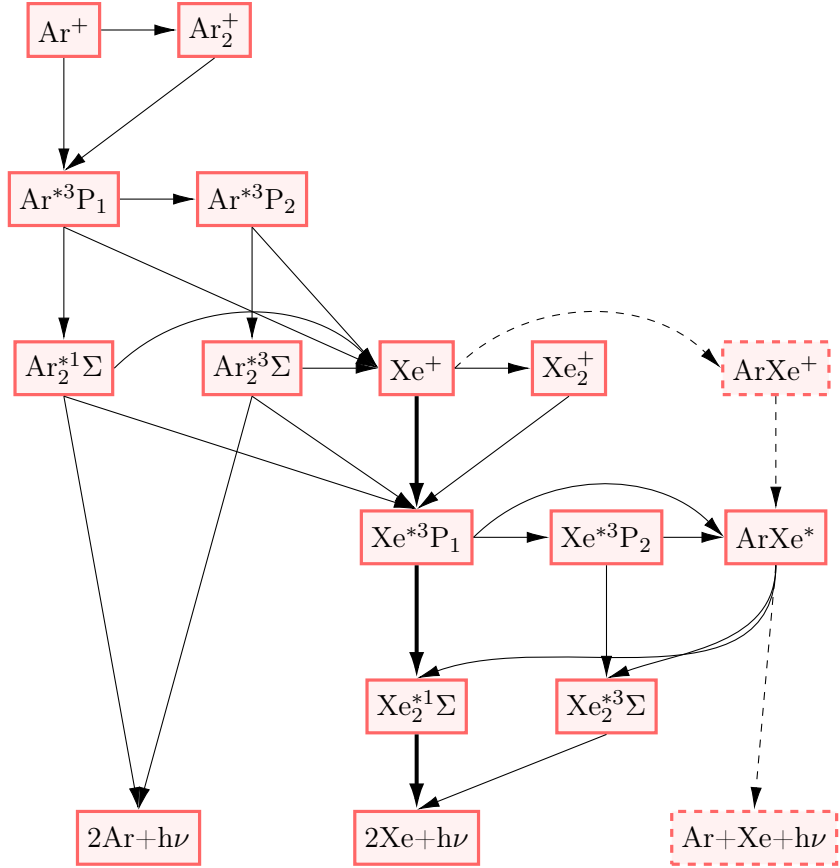

In pure liquid argon, incoming particles can produce argon atomic excitations that quickly relax to $^{3}\mathrm{P}_{1}$ and $^{3}\mathrm{P}_{2}$ states. 
The $^{3}\mathrm{P}_{1}$ and $^{3}\mathrm{P}_{2}$ atoms then form homonuclear diatomic molecular excimers: $^{1}\Sigma^{+}_{\mu}$ and $^{3}\Sigma^{+}_{\mu}$, respectively, through a three-body process, which happens on a time scale of picoseconds in liquid phase~\cite{Physica82C197619-26}. 
Photon emissions from the decays of the $^{1}\Sigma_{\mu}^{+}$ and the $^{3}\Sigma_{\mu}^{+}$ states to the repulsive ground state $^{1}\Sigma_{g}^{+}$ are identified as the fast and the slow component of liquid argon scintillation light, respectively. 
The lifetime of the $^{3}\Sigma_{\mu}^{+}$ state is longer than that of the $^{1}\Sigma_{\mu}^{+}$ state, because the direct decay from $^{3}\Sigma_{\mu}^{+}$ to $^{1}\Sigma_{g}^{+}$ is forbidden~\cite{kubota1978triplet}.
It is also possible to produce argon ions through ionization resulting from incoming particles. 
The ionized argon atom ($\Ar^{+}$) collides with another argon atom to form $\Ar_{2}^{+}$ in a short time, then recombines with electrons and relaxes to the $^{3}\mathrm{P}_{1}$ and  $^{3}\mathrm{P}_{2}$ states. 
The rest of the processes are the same as those described above.
Detailed recombination, relaxation, and energy transfer processes are shown in figure~\ref{fig:process}. 
In figure~\ref{fig:process} and the rest of this article, the excited molecular state $^{1}\Sigma_{\mu}^{+}$ and $^{3}\Sigma_{\mu}^{+}$ are abbreviated as $^{1}\Sigma$ and $^{3}\Sigma$, respectively.

When a xenon dopant is present, $\Ar^{*}$ and $\Ar_{2}^{*}$ can ionize xenon atoms through the Penning ionization process before forming excimers~\cite{miller1970PenningIonization}. 
The $\Ar_{2}^{*}$ energy level is a good match to the  $^3\mathrm{P}_1$ state of $\Xe^{*}$, which results in an effective energy transfer to xenon atomic excitation. 
$\Xe^{*}$ then goes through a relaxation process that is similar to that for $\Ar^{*}$, with an additional pathway with the formation of the heteromolecular excimer $\Ar\Xe^{*}$. 
$\Ar\Xe^{*}$ radiatively decays, emitting a photon of wavelength \SI{150}{nm}, which has been observed in a specific gas phase setup reported in~\cite{nowak1985heteronuclear}. 
However, due to the relatively low binding energy of $\Ar\Xe^{*}$ compared to the binding energy of $\Xe_{2}^{*}$, this photon emission channel is expected to be negligible at high xenon concentrations in liquid~\cite{kolts1978decayInAr, ermler1993abInitioXe2Energy}. 
In addition, $\Ar\Xe^{*}$ might be responsible for vibrational relaxation from high energy states~\cite{brodmann1978xenon}.
The formation of a hump structure in the scintillation light time profile of the mixture was attributed to the existence of $\Ar\Xe^{*}$ by S.~Kubota et al.~\cite{kubota1993suppression}. 
On the other hand, an excited argon atom and a ground state xenon atom will not form the excimer $\Ar^{*}\Xe$ due to a repulsive potential curve that crosses the binding curves \cite{nowak1985heteronuclear}.  
Thus, $\Ar^{*}$ is not able to transfer energy to $\Xe^{*}$ directly through the formation of $\Ar^{*}\Xe$.

The model of the energy dissipation process is represented by equation~\ref{eq:process}, where $N_{0}$ is the number of argon atoms in excited states and $N_\mathrm{A}$ is the number of A.
$k_{\mathrm{A,B}}$ denotes the reaction constant for the process $\mathrm{A}\to\mathrm{B}$, whose rate is proportional to the xenon concentration $C_{\mathrm{Xe}}$. 
$l_{\mathrm{A,B}}$ denotes the rate that is independent of xenon concentration. $\tau_\mathrm{A}$ denotes the lifetime of A. 
$p_\Ar$($p_\Xe$) denotes the probability of forming singlet state argon (xenon) excimers. 
$R$ is the rate of photon emission.
This model helps us to quantitatively understand the evolution of the scintillation light time profile with different xenon concentrations. 
It also helps us to find the minimum xenon concentration that is able to completely shift the slow component of liquid argon scintillation to the prompt.

\begin{equation}
\label{eq:process}
    \begin{aligned}
        \frac{dN_0}{dt} &= - l_{\Ar^*,\Ar_2^*}N_0 -k_{\Ar^*,\Xe^*}C_{\Xe}N_0 \\   
        \frac{dN_{\Ar_2^{*3}\Sigma}}{dt}    &= (1-p_{\Ar}) l_{\Ar^*,\Ar_2^*}N_0 - \left(\frac{1}{\tau_{\Ar_2^{*3}\Sigma}} + k_{\Ar_2^{*3}\Sigma,\Xe^*} C_{\Xe}\right) N_{\Ar_{2}^{*3}\Sigma} \\
        \frac{dN_{\Ar_2^{*1}\Sigma}}{dt}    &= p_{\Ar} l_{\Ar^*,\Ar_2^*} N_0 - \left(\frac{1}{\tau_{\Ar_2^{*1}\Sigma}} +k_{\Ar_2^{*}{}^1\Sigma, \Xe^*}C_{\Xe}\right) N_{\Ar_2^{*1}\Sigma}\\
        \frac{dN_{\Xe^*}}{dt}     &= k_{\Ar_2^{*3}\Sigma,\Xe^*}C_{\Xe} N_{\Ar_2^{*3}\Sigma} + k_{\Ar_2^{*1}\Sigma,\Xe^*}C_{\Xe} N_{\Ar_2^{*1}\Sigma} + k_{\Ar^*,\Xe^*}C_{\Xe}N_0 \\
        &- \left(k_{\Xe^*,\Xe_2^{*1}\Sigma} + k_{\Xe^*,\Xe_2^{*3}\Sigma}\right) C_{\Xe} N_{\Xe^*} - l_{\Xe^*,\Ar\Xe^*} N_{\Xe^*}\\
        \frac{dN_{\Ar\Xe^*}}{dt}     &= l_{\Xe^*,\Ar\Xe^*} N_{\Xe^*} - k_{\Ar\Xe^*,\Xe_2^*} C_{\Xe} N_{\Ar\Xe^*}\\
        \frac{dN_{\Xe_2^{*3}\Sigma}}{dt}      &= (1-p_{\Xe})k_{\Ar\Xe^*,\Xe_2^*} C_{\Xe} N_{\Ar\Xe^*} +  k_{\Xe^*,\Xe_2^{*3}\Sigma}C_{\Xe}N_{\Xe^*} - \frac{N_{\Xe_2^{*3}\Sigma}}{\tau_{\Xe_2^{*3}\Sigma}}\\
        \frac{dN_{\Xe_2^{*1}\Sigma}}{dt}      &= p_{\Xe} k_{\Ar\Xe^*,\Xe_2^*} C_{\Xe} N_{\Ar\Xe^*} +  k_{\Xe^*,\Xe_2^{*1}\Sigma}C_{\Xe}N_{\Xe^*} - \frac{N_{\Xe_2^{*1}\Sigma}}{\tau_{\Xe_2^{*1}\Sigma}}\\
        R_{\SI{128}{nm}}  &= \frac{N_{\Ar_2^{*1}\Sigma}}{\tau_{\Ar_2^{*1}\Sigma}} + \frac{N_{\Ar_2^{*3}\Sigma}}{\tau_{\Ar_2^{*3}\Sigma}}\\ 
        R_{\SI{178}{nm}}  &= \frac{N_{\Xe_2^{*1}\Sigma}}{\tau_{\Xe_2^{*1}\Sigma}} + \frac{N_{\Xe_2^{*3}\Sigma}}{\tau_{\Xe_2^{*3}\Sigma}} \\  
    \end{aligned}
\end{equation}

When TPB is present, photons emitted from argon and xenon are absorbed by the TPB and then re-emitted. 
Therefore, an additional equation~\ref{eq:process_TPB} which describes the TPB fluorescence process has to be considered. 
In this equation, $N_{\mathrm{TPB}^*}$ represents the number of TPB molecules that are in excited states after absorbing VUV photons.
$E_{\SI{128}{nm}}$ ($E_{\SI{178}{nm}}$) represents the wavelength shift efficiency (WLSE) of TPB at \SI{128}{nm} (\SI{178}{nm}). 

\begin{equation}
\label{eq:process_TPB}
    \begin{aligned}
    \frac{dN_{\mathrm{TPB}^*}}{dt} &=E_{\SI{128}{nm}} R_{\SI{128}{nm}} + E_{\SI{178}{nm}}R_{\SI{178}{nm}} - \frac{N_{\mathrm{TPB}^*}}{\tau_{\mathrm{TPB}}} \\
    R_{\SI{420}{nm}} &= \frac{N_{\mathrm{TPB}^*}}{\tau_{\mathrm{TPB}}}
    \end{aligned}
\end{equation}

This process is equivalent to convolving the scintillation light time profile with the exponential response contributed by the TPB, which delays the rising edge of the scintillation light pulse shape by around $\tau_\mathrm{TPB}$. 
Therefore, to compete with TPB in fast timing applications, xenon concentration should be high enough that the sum of the reaction time constants (the reciprocals of the reaction rates) along the pathway is in the sub-nanosecond range.

\section{Experimental apparatus}
\begin{figure}[ht]
\centering 
\subfigure[Test setup]{ \label{fig:system:a} 
\includegraphics[width=8cm]{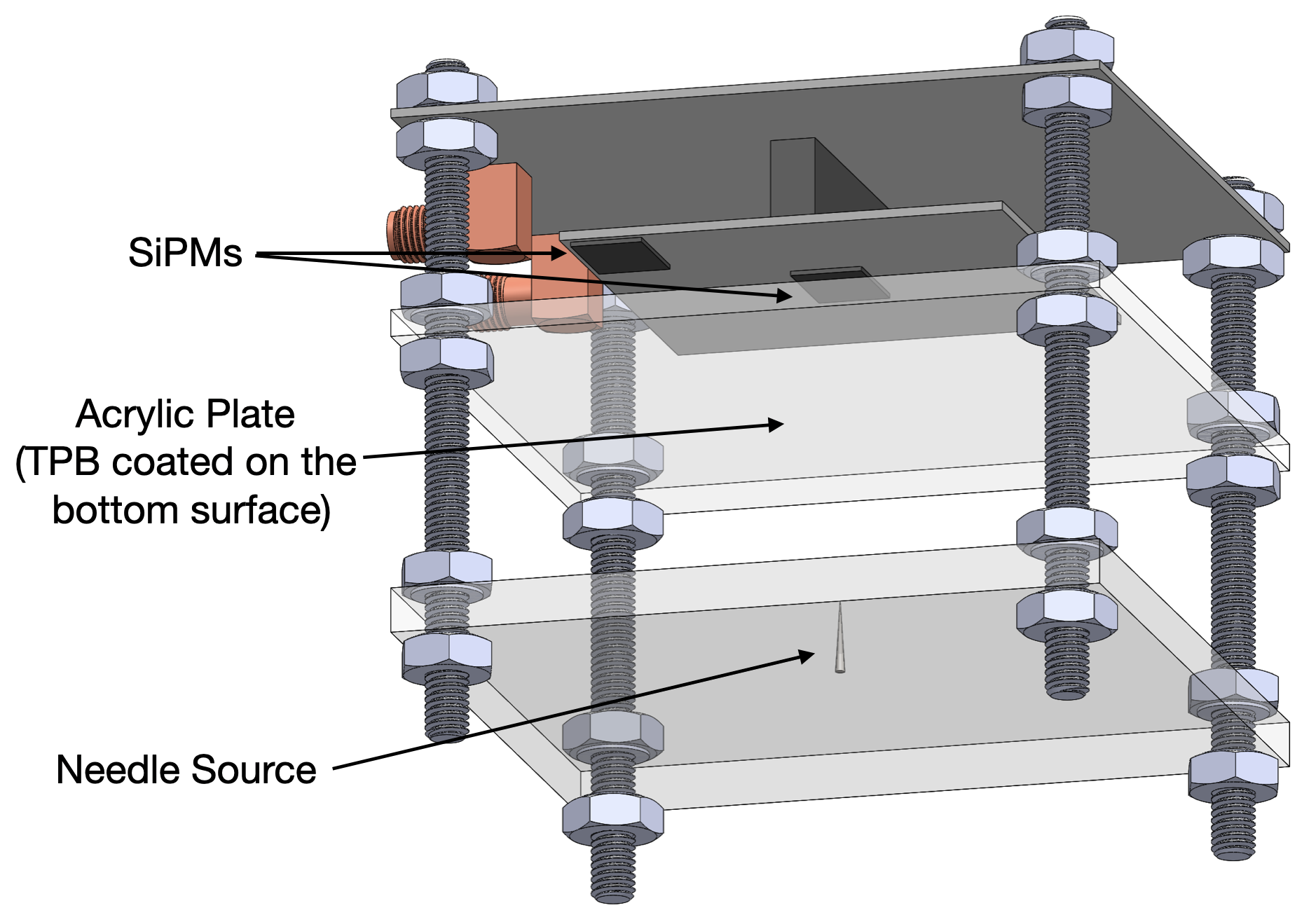}} 
\hspace{0.1in} 
\subfigure[System P\&ID]{ \label{fig:system:b} 
\includegraphics[width=6cm]{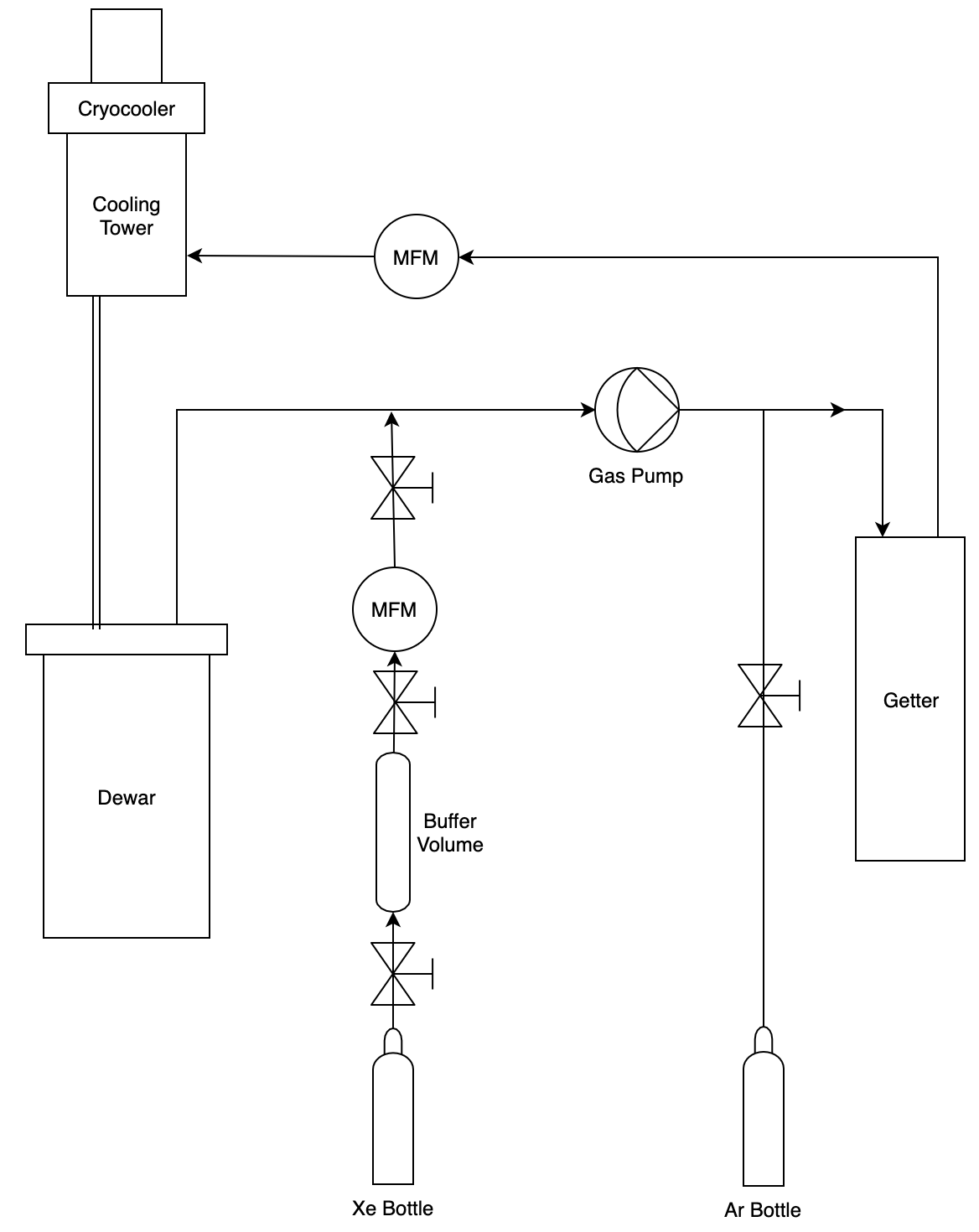}} 
\caption{$\emph{Left}$: 3D view of the test setup.  From the top to the bottom is the cold front-end electronics, two SiPM channels, an acrylic plate coated with TPB on the bottom surface and the needle source fixed at the center of a high density polyethylene plate; $\emph{Right}$: piping and instrumentation diagram (P\&ID) of the test system, which consists of a cryocooler, a cooling tower, a stainless steel dewar, a gas circulation pump, and a gas getter. The incoming argon gas is injected to the system directly from the gas bottle; and the incoming xenon gas is first injected into a buffer volume, and then slowly drawn into the system.} 
\label{fig:system} 
\end{figure}

Figure~\ref{fig:system:a} shows the test setup for this study. 
Two SiPMs are mounted on a printed circuit board as two different readout channels. 
Each readout channel is driven by a cold pre-amplifier, which is attached on the backside of the SiPM.
A TPB-coated acrylic plate is located about \SI{5}{mm} below the front surface of the SiPMs, while the TPB-coated side is facing downward.
About \SI{2.5}{cm} below the front surface of the SiPM, there is a needle source mounted at the center of a reflective high density polyethylene plate.
During the test, two needle sources, $^{210}\mathrm{Po}$ and $^{90}\mathrm{Sr}$, are deployed separately in different runs.
The central SiPM is used for the pulse shape analysis, due to its high photon collection efficiency.
The SiPM at the corner, where the photon rate is lower,  is used as a reference.

This setup is placed in a double-wall stainless steel dewar, which is connected to the cryogenic system.
Figure~\ref{fig:system:b} shows the plumbing and instrumentation of the system.
The incoming argon gas from the gas bottle is pushed into a hot getter for purification before entering the cooling tower for condensation by the cryocooler.
Liquefied argon flows  through a double-wall-isolated transfer line into the stainless steel dewar, where the test cell is situated.   The vaporized argon gas is pushed back into the getter by the gas recirculation pump.
To minimize the risk of xenon-ice formation during the doping process, the xenon gas is initially injected into a buffer volume upto \SI{100}{kPag} in a single shot, and then slowly drawn into the system and mixed with the re-circulating argon gas.

At the beginning of the test, pure argon is condensed and fills the dewar to the desired level (3~liters of liquid in total), such that the test apparatus is fully immersed in the liquid. 
Once full, the system is switched into a circulation mode to purity the argon using a SAES Micro-Torr gas getter, a step that takes more than 72~hours with a circulation speed of \SI{8}{stdL/min}.
After the purification process, calibration data are collected to characterize the performance of the SiPMs.
Once the SiPM characterization is done, xenon is injected in batches into the system through a small buffer chamber.
Each batch increases the total xenon concentration in the system by \SI{20}{ppm}, which is quantified by the volume and the pressure of the buffer chamber.
Xenon in the buffer chamber is slowly released into the system over 2~minutes, followed by a 5~minutes system circulation before the next injection.  
After the desired amount of xenon is injected, data is taken every 12 hours to monitor changes in the pulse shape from the scintillation light.
Once the pulse shape remains stable for three successive runs, the injected xenon is considered to be uniformly distributed. This process usually takes 36 to 48 hours, and the last set of data is accepted for the final analysis.

The SiPMs used in this work are the low after-pulsing rate FBK NUV sensitive SiPMs, which are specially designed for liquid argon detectors~\cite{gola2019nuv}. 
The SiPM is optimized to reach peak photon detection efficiency (PDE) to near ultraviolet light, and it is also sensitive to the xenon scintillation light. 
The pre-amplifier used in this work is described in~\cite{DIncecco2018preamp}. 

Gas in the system has been liquefied  three times with different configurations: $^{210}\mathrm{Po}~\alpha$ with TPB, $^{210}\mathrm{Po}~\alpha$ without TPB and $^{90}\mathrm{Sr}~\beta$ without TPB. 
Testing with different particles is done to obtain information on particle identification with PSD in the mixture~\cite{Mckinsey2008ArPSD}.
Pulse shapes with different fast-to-slow-component ratios serve to further constrain the parameters of the model used to fit the data.   Since the SiPMs have low quantum efficiency for the \SI{128}{nm} light, when no TPB is coated in the system, the SiPMs basically only detect the fraction of light emitted by xenon, whereas with the TPB coating, 
argon and xenon emissions are detected with similar efficiency.  
Thus, a comparison between configurations with and without TPB will provide spectral information.
Data in each configuration is acquired with two different SiPM biases, \SI{3.5}{V} and \SI{5.5}{V} over-voltage. 
The SiPMs are operated in the regime of low to medium over-voltages, to reduce the correlated noise. 
Each configuration has xenon doping concentrations up to \SI{1600}{ppm} by mole fraction.

\section{Analysis and results}
\subsection{Pulse identification}

The acquired waveform is processed by a precise pulse finding algorithm that includes four steps: pole-zero filtering, preliminary peak finding, baseline finding, and fine pulse start identification.

\begin{figure}
    \centering
    \includegraphics[width = \textwidth]{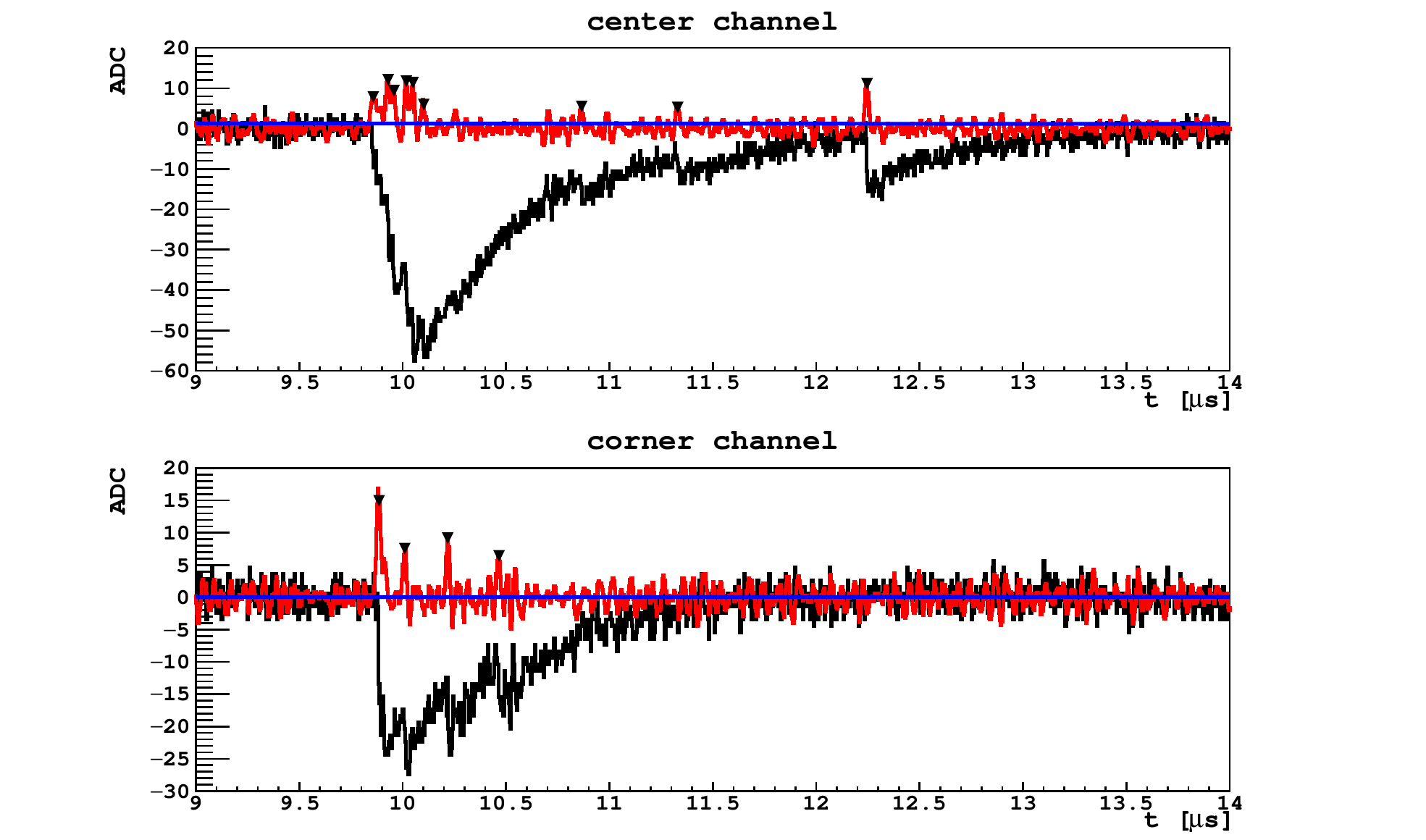}
    \caption{A randomly selected event as an example of the pulse finding algorithm. Top and bottom panels are for the center and corner channels, respectively. The black waveform is the original with baseline subtraction. The blue line is the reconstructed baseline. The red waveform is the fast component after applying the pole-zero filter to the black. Inverted black triangles indicate the identified photo-electron peaks.}
    \label{fig:wfexample}
\end{figure}

A digital pole-zero filter is applied to remove the slow decay tail of the SiPM response.
A processed waveform is shown as a red line in figure~\ref{fig:wfexample}.
A peak-finding algorithm from the \texttt{ROOT TSpectrum} class is then applied to identify the arrival time 
of individual photons.
Any unresolved photons that arrive close in time are identified as a single hit with large amplitude.
Once the peaks are found, the baseline is calculated from the average of the raw waveform in photon-peak-excluded windows.   The processed baseline is shown as a blue line in figure~\ref{fig:wfexample}.
Finally, for each registered peak that is not on the tail of a previous peak, the zero-crossing time of a linear fit to the first four samples on the rising edge of the raw waveform is defined as the pulse start time.

Based on this pulse identification procedure, the data from all the three system configurations are analyzed.   The results are shown in the rest of this section.

\subsection{System stability}

Since the duration of the full set of tests was long and the system went through thermal cycles several times to change the test configuration, the system stability was closely monitored throughout the series of tests.
In terms of the cryogenic system, the temperature of liquid, which was measured by PT-100 temperature sensors, was controlled within the range \SI{85.9\pm0.3}{K}.  The pressure was controlled within the range \SI{13.7\pm0.1}{psi}.

\begin{figure}
    \centering
    \includegraphics[width = \textwidth]{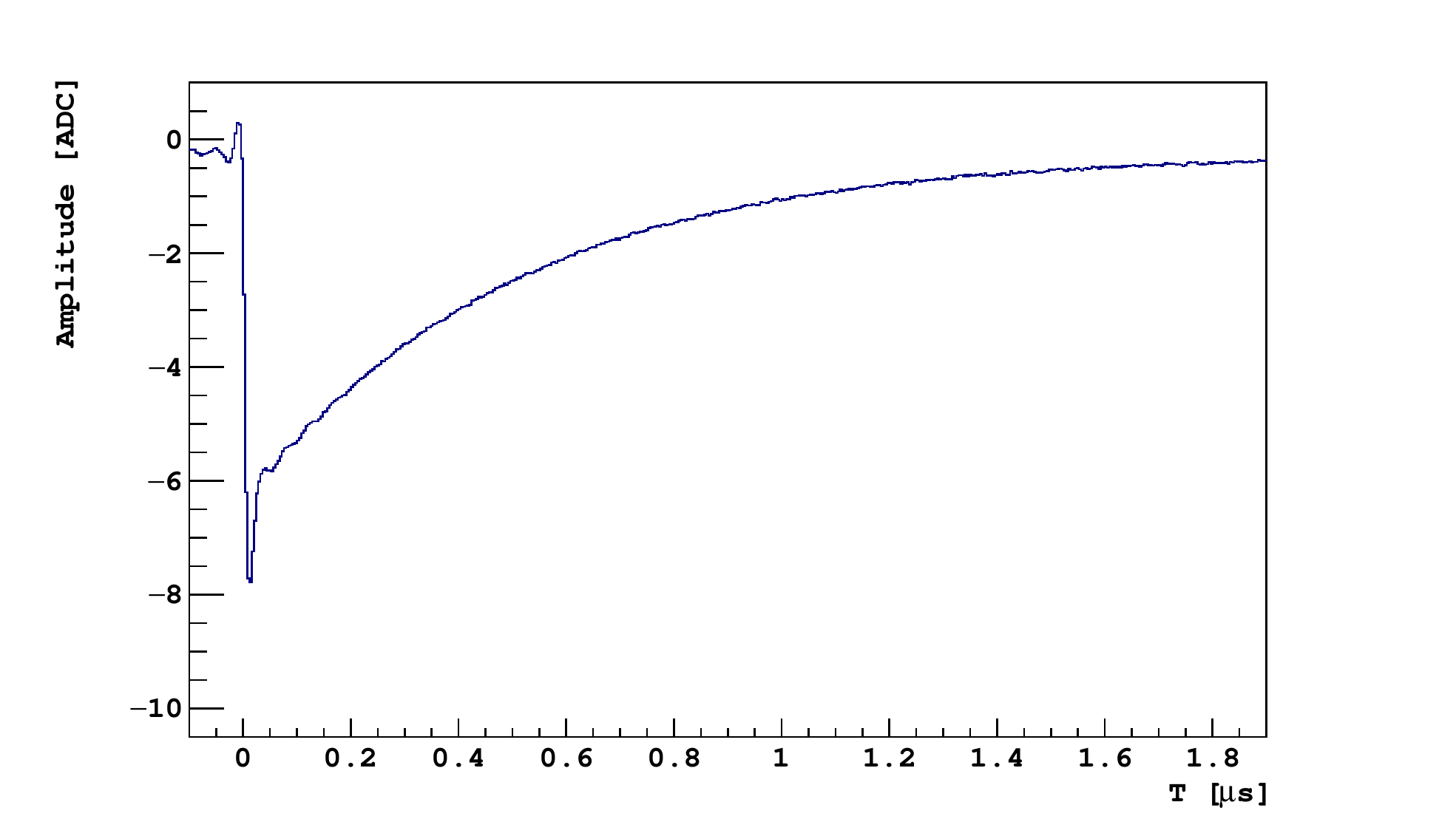}
    \caption{Example of the SPE pulse at \SI{3.5}{V} over-voltage.}
    \label{fig:spepulse}
\end{figure}

In terms of the photoelectronics, the center SiPM is calibrated after each thermal cycle and each doping procedure. 
The SiPM single photo-electron (SPE) gain is stable with the two over-voltages during the entire test. 
Measurement shows that the fluctuation in SPE gain is \SI{0.5}{\percent} at \SI{3.5}{V} over-voltage and \SI{0.6}{\percent} at \SI{5.5}{V} over-voltage in RMS. 
The average SPE pulse shape from each run is also highly stable. 
No noticeable deformation of the SPE pulse shape is found throughout the entire test. 
An example SPE pulse is shown in figure~\ref{fig:spepulse}.

\begin{figure}
    \centering
    \subfigure[]{\label{fig:ap:3p5v}
    \includegraphics[width = \textwidth,  trim=0.8cm 0.2cm 1.2cm 0.5cm, clip=true]{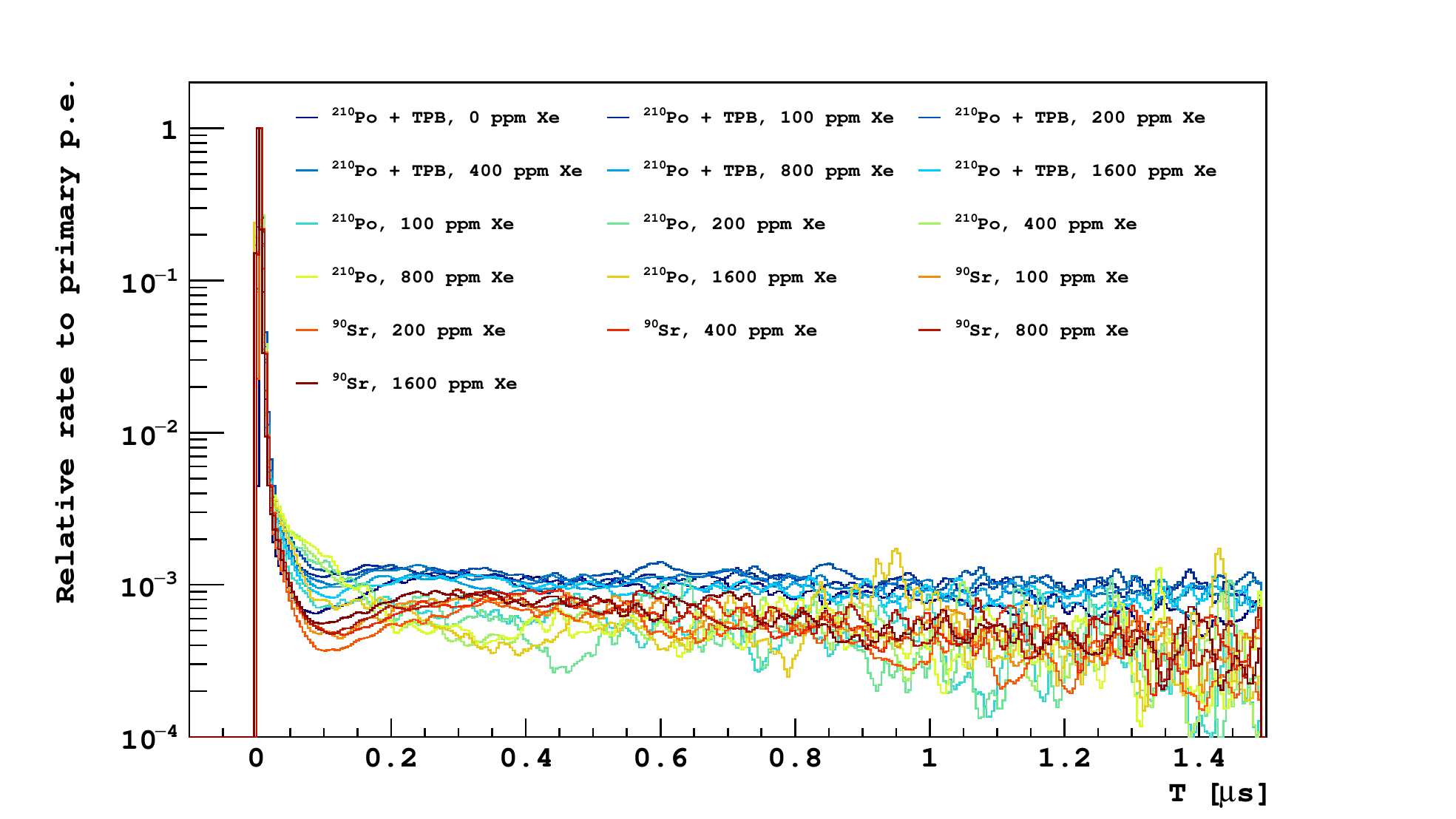}}
    \subfigure[]{\label{fig:ap:5p5v}
    \includegraphics[width = \textwidth,  trim=0.8cm 0.2cm 1.2cm 0.5cm, clip=true]{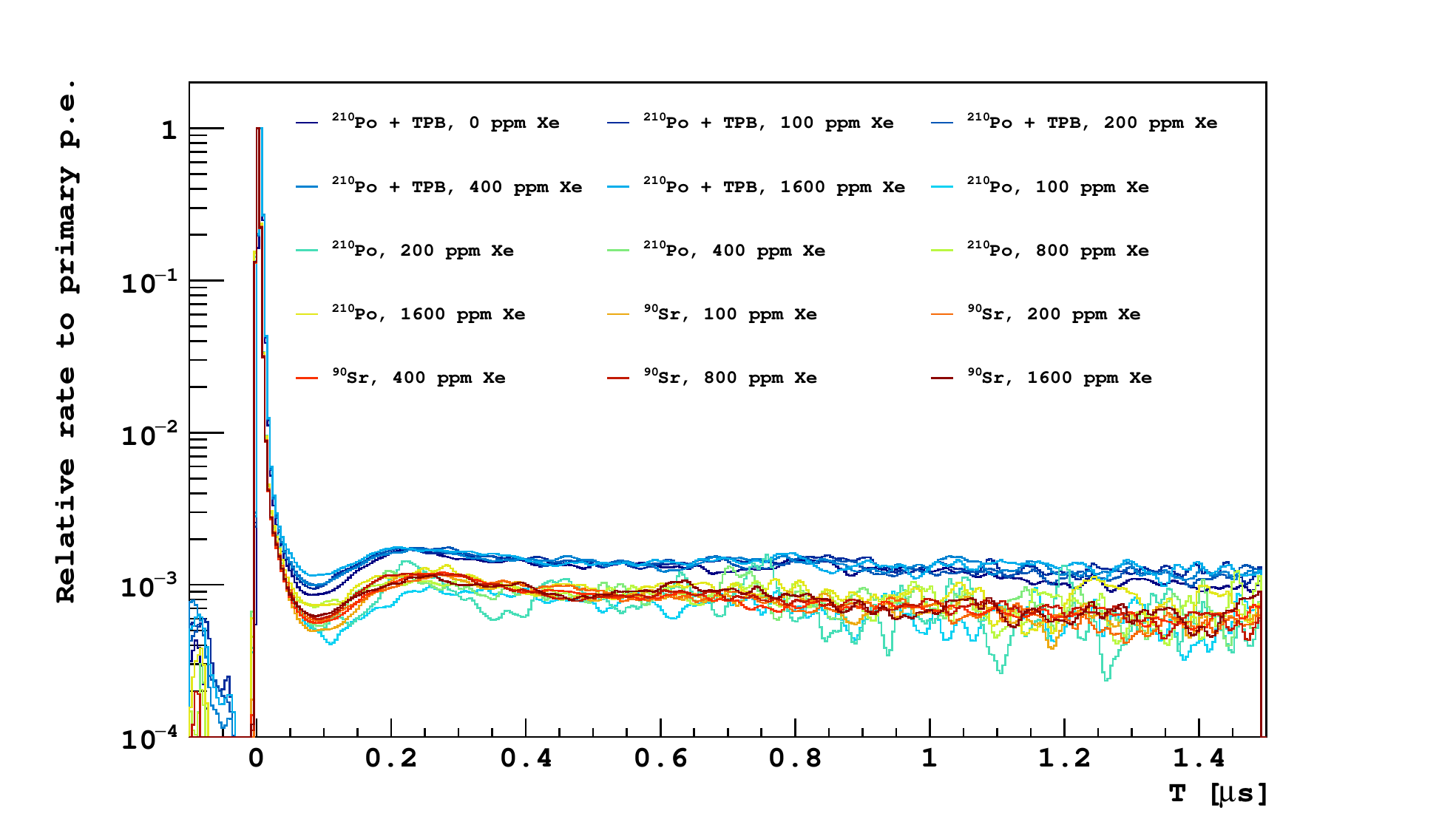}}
    \caption{Distribution of the time difference between successive SPEs outside the signal region of the waveform. The peak at $T=0$ is the primary photo-electron. The broad distribution following the peak are the correlated noise and random coincidence. \emph{Top}: SiPM operated under \SI{3.5}{V} over-voltage. \emph{Bottom}: SiPM operated under \SI{5.5}{V} over-voltage.}
    \label{fig:ap}
\end{figure}

Since the correlated noise could alter the pulse shape when its rate is sufficiently high, the correlated noise rate is also estimated for each run.
Time correlations between SPE pulses are searched for in the scintillation-signal-excluded windows. 
These SPE pulses include dark counts, faint scintillation light from environmental background, and leakage of the slow component of the source events. 
Figure~\ref{fig:ap} shows the distribution of the time difference between the primary photo-electron and  successive photons, with the primary photo-electron taken to be at $T=0$.
Each distribution includes a constant pedestal from random coincidences.
The non-flat structures reveal the existence of correlated noise.
Distributions from all the runs show good consistency and the correlated noise rate is around one thousandth of the primary. 
This proves that we can safely neglect the correlated noise in the pulse shape analysis.

\subsection{Pulse shape}

In order to understand the changes in the time profile of scintillation light as a function of xenon concentration, the pulse shapes of events are carefully studied.

Since the acquired waveform is a convolution of the scintillation light time profile and the SiPM response (and in some configurations, the TPB time response), a deconvolution process has been applied to the waveform to disentangle the scintillation light time profile from the detector response. 
As the first step, the average waveform of each run is calculated.  Subsequent to that,  the average pulse shape of the single photoelectron signal is acquired to represent the single photon response (SER) of the SiPM.
The deconvolution algorithm from \texttt{ROOT}'s \texttt{TSpectrum} is applied to remove the effect of the SER in the acquired waveform.
After this process, the waveform is ready to be fitted to the model described in equation~\ref{eq:process} and equation~\ref{eq:process_TPB}.

\begin{figure}
    \centering
    \subfigure[]{\label{fig:f90:example}
    \includegraphics[width = 6.9cm, trim=0.8cm 0.cm 2.2cm 0.cm, clip=true]{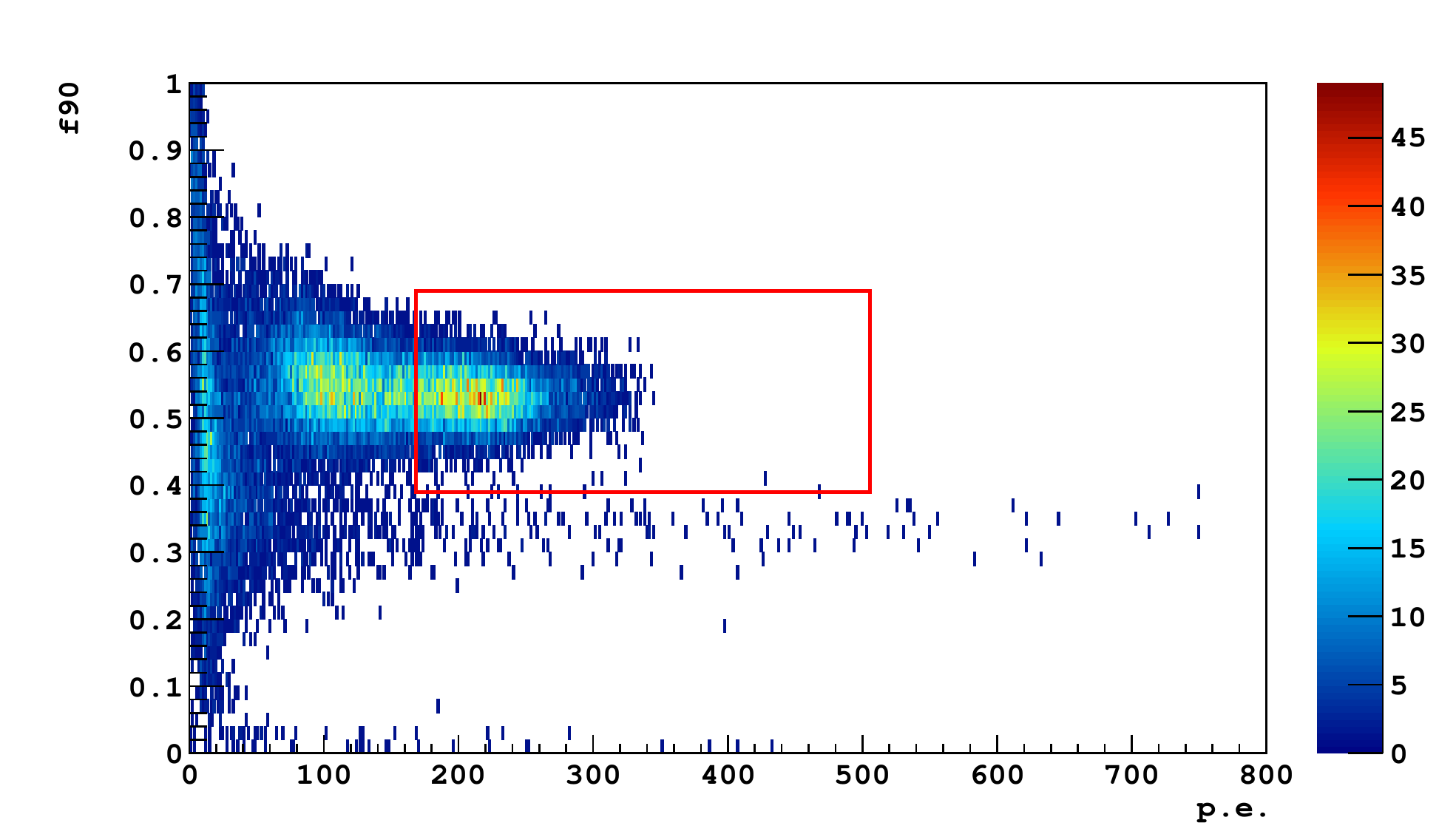}}
    \hspace{0.1in}
    \subfigure[]{\label{fig:f90:trend}
    \includegraphics[width = 7.4cm, trim=0.8cm 0.cm 2.6cm 0.cm, clip=true]{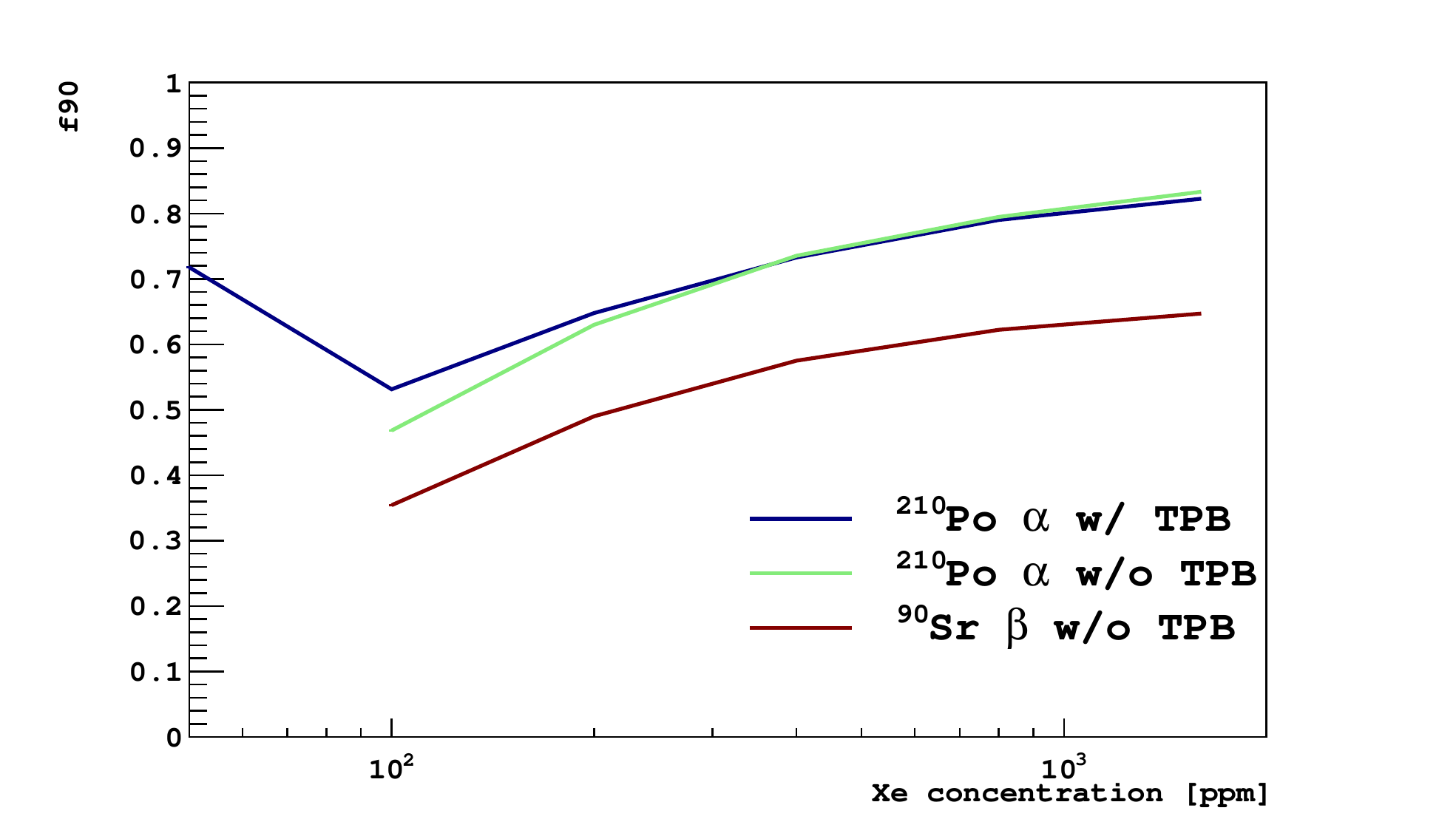}}
    \caption{F90 of the pulses. \emph{Left}: f90 plot of $^{210}\mathrm{Po}~\alpha$ data with \SI{100}{ppm} xenon and TPB. The red box indicates the acceptance region of events for the average waveform. The lower band consists of background from the environment. \emph{Right}: The trend of f90 along with different xenon concentrations. The 2 curves with $\alpha$ source are consistent and higher than the curve of $\beta$ source. The point on the y-axis shows the f90 in pure liquid argon.}
    \label{fig:f90}
\end{figure}

Basic pulse quality cuts are applied to remove saturated pulses and dark counts.
The pulse shape cut is applied to select $\alpha$ or $\beta$ events according to the type of source, with the parameter called f90, which is defined as the light fraction in the first \SI{90}{ns} of the pulse.
F90 is expected to change along with the xenon concentration.
Figure~\ref{fig:f90:example} shows an example of the f90 distribution versus energy of events in a run with a \SI{100}{ppm} xenon concentration, TPB, and the $^{210}\mathrm{Po}$ source. 
The accepted region in the f90-energy parameter space is defined as a box around the shoulder of the spectrum run by run, shown as the red box in figure~\ref{fig:f90:example}.
The trends of f90 versus xenon concentration of $\alpha$ and $\beta$ events are shown in figure~\ref{fig:f90:trend}. 
The $^{210}\mathrm{Po}~\alpha$ with TPB line indicates that from pure liquid argon to \SI{100}{ppm} xenon concentration, f90 exhibits a decreasing trend.  This trend changes to an increase as more xenon is added.  The decrease in the range of low xenon concentration is due to the fact that the excitation energy has been significantly shifted to xenon atoms, but the formation of $\Xe^{*}_{2}$ is slow.
This is consistent with the phenomenon that a hump has been observed in the deconvolved waveform at \SI{100}{ppm} xenon concentration.   
The other two configurations ($^{210}\mathrm{Po}~\alpha$ without TPB and $^{90}\mathrm{Sr}~\beta$ without TPB) only show the trend from \SI{100}{ppm} xenon concentration to \SI{1600}{ppm}, due to the lack of PDE to argon scintillation.

Combined fits to waveforms from all data sets are carried out with two assumptions.  
The first is that the events taken under the same xenon concentration with and without TPB have the same reaction rates.   This assumption is valid owing to the reliability and stability of the cryogenic system.
The second assumption is that the rates of xenon-involved reactions are proportional to the xenon concentration. 
It is valid since the overall xenon concentration is low, so reactions that involve two free xenon 
atoms can be neglected.   
The fitting range is chosen to be from the pulse start to \SI{0.25}{\us}, since after around \SI{0.3}{\us} the light decays to a thousandth of the peak intensity and correlated noise caused by prompt light becomes dominant, as shown in figure~\ref{fig:ap}. 
This fitting window also emphasizes the fast component, which is the most interesting part of this study. 
Due to the timing resolution of the system, pulses are not perfectly aligned for the average, so a \SI{2}{ns} RMS smearing is introduced to improve the fitting.   
Some parameters are fixed to values found in the literature (listed in table~\ref{tab:fixpars}).
Some detailed discussion of the parameters in the list can be found in appendix~\ref{freepara}.  

\begin{table}
    \centering
    \begin{tabular}{c|c|c}
         \hline
         Parameter      & Fixed value   & Comment                          \\
         \hline 
         $l_{\Ar^*,\Ar_2^*}$    & \SI{1.66e5}{\us^{-1}}    & Rate of $\Ar^{*}$ trapping by $\Ar$~\cite{martin1971excition}.\\
         $k_{\Ar^*,\Xe^*}$       & \SI{7e-9}{cm^3/s}    & \makecell{Rate coefficient of $\Xe$ to be excited or \\ionized by $\Ar$ \cite{hitachi1984excitonRGC,kubota1976XeIonLAr}.}\\
         $k_{\Ar_{2}^{*}{}^{3}\Sigma,\Xe^{*}}$  & \SI{1e-11}{cm^3/s} & \makecell{ Rate coefficient of $\Xe$ to be excited \\by $\Ar_2^*{}^3\Sigma$ \cite{kubota1982liquid,kubota1993suppression,wahl2014pulse}.} \\
         $k_{\Xe^{*},\Xe_2^*{}^1\Sigma}$  & \SI{4.4e-9}{cm^3/s} & Scaled by density from gas phase value \cite{gleason1977ArXeETransfer}.\\
         $k_{\Xe^{*},\Xe_2^*{}^3\Sigma}$  & \SI{4.5e-10}{cm^3/s} & Scaled by density from gas phase value \cite{gleason1977ArXeETransfer}.\\
         $\tau_{\Ar,s}$  & \SI{7}{ns}    & Decay time of $\Ar_2^{*}{}^{1}\Sigma$ to \SI{128}{nm} photon \cite{hitachi1983IonDensityPSD}.\\
         $\tau_{\Ar,l}$  & \SI{1.6}{\us} & Decay time of $\Ar_2^{*}{}^{3}\Sigma$ to \SI{128}{nm} photon \cite{hitachi1983IonDensityPSD}.\\
         $\tau_{\Xe,s}$  & \SI{4.3}{ns}  & Decay time of $\Xe_2^{*}{}^{1}\Sigma$ to \SI{178}{nm} photon \cite{hitachi1983IonDensityPSD}.\\
         $p_{\Ar \beta}$    & 0.3        & Singlet ratio of $\Ar_2^{*}$ from $\beta$ scattering \cite{wahl2014pulse}.\\
         $p_{\Ar \alpha}$    & 0.6    & Singlet ratio of $\Ar_2^{*}$ from $\alpha$ scattering \cite{wahl2014pulse}.\\
         $\tau_{TPB}$   & \SI{1}{ns} &  TPB light lifetime.\\
         WLSE & 0.48 & \makecell{TPB wavelength shift efficiency\\at \SI{128}{nm} and \SI{178}{nm}~\cite{benson2018tpb}.}\\
         $\mathrm{PDE}_{\SI{128}{nm}}/ \mathrm{PDE}_{\SI{420}{nm}}$ & 0.029 & \makecell{Estimated based on maximum\\ pulse amplitude in pure LAr.}\\
         \hline
    \end{tabular}
    \caption{Fixed parameters. The reference value of $k_{\Xe^{*},\Xe_{2}^{*}}$ is calculated with liquid argon density assuming the same reaction rate in gas phase \cite{gleason1977ArXeETransfer}.}
    \label{tab:fixpars}
\end{table}

\begin{figure}[ht]
\centering 
\subfigure[]{ \label{fig:ps:210Po30p5v} 
\includegraphics[width=7.3cm, trim=0.8cm 0.2cm 2.1cm 0.5cm, clip=true]{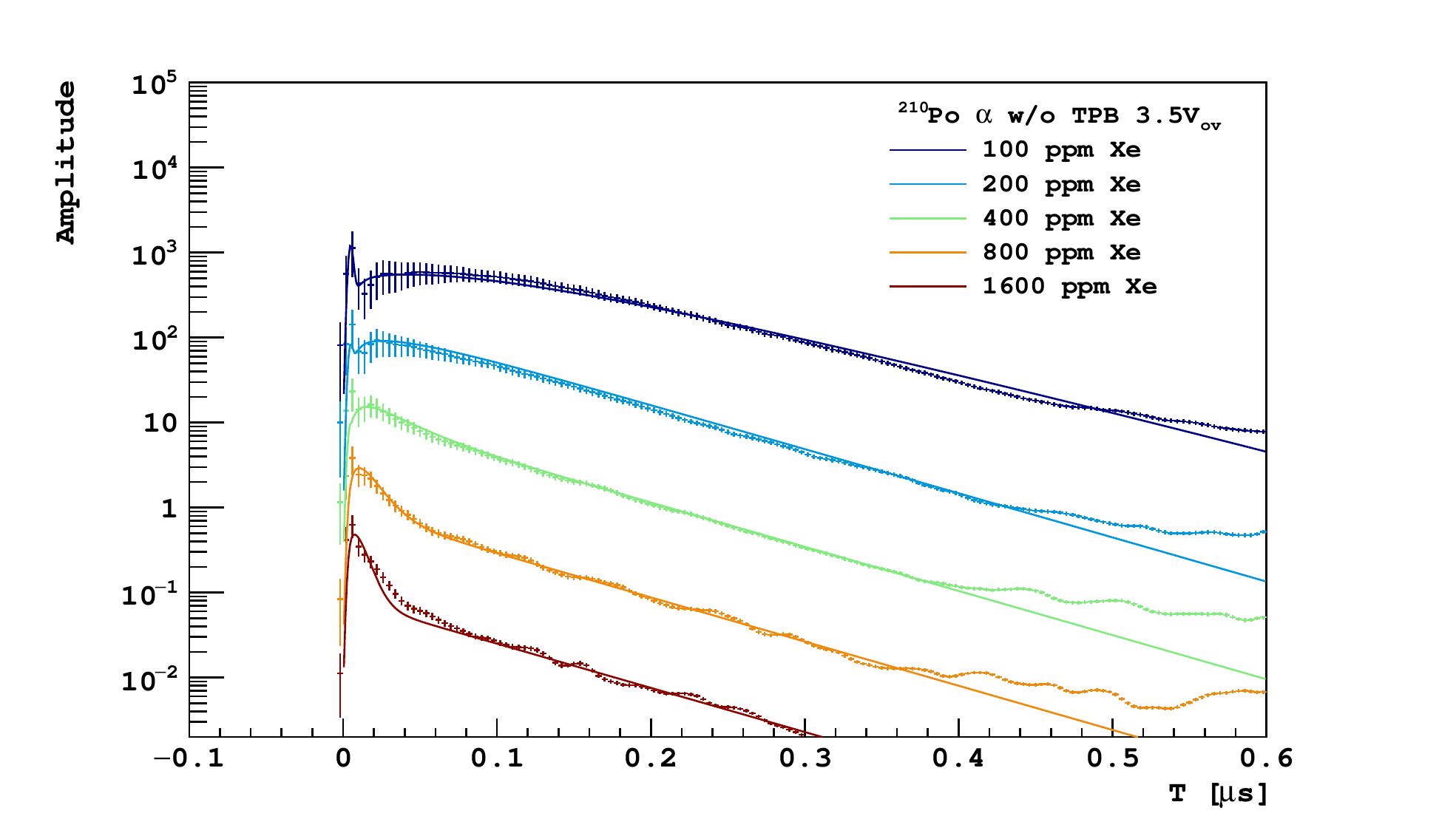}} 
\subfigure[]{ \label{fig:ps:210Po32p5v} 
\includegraphics[width=7.3cm, trim=0.8cm 0.2cm 2.1cm 0.5cm, clip=true]{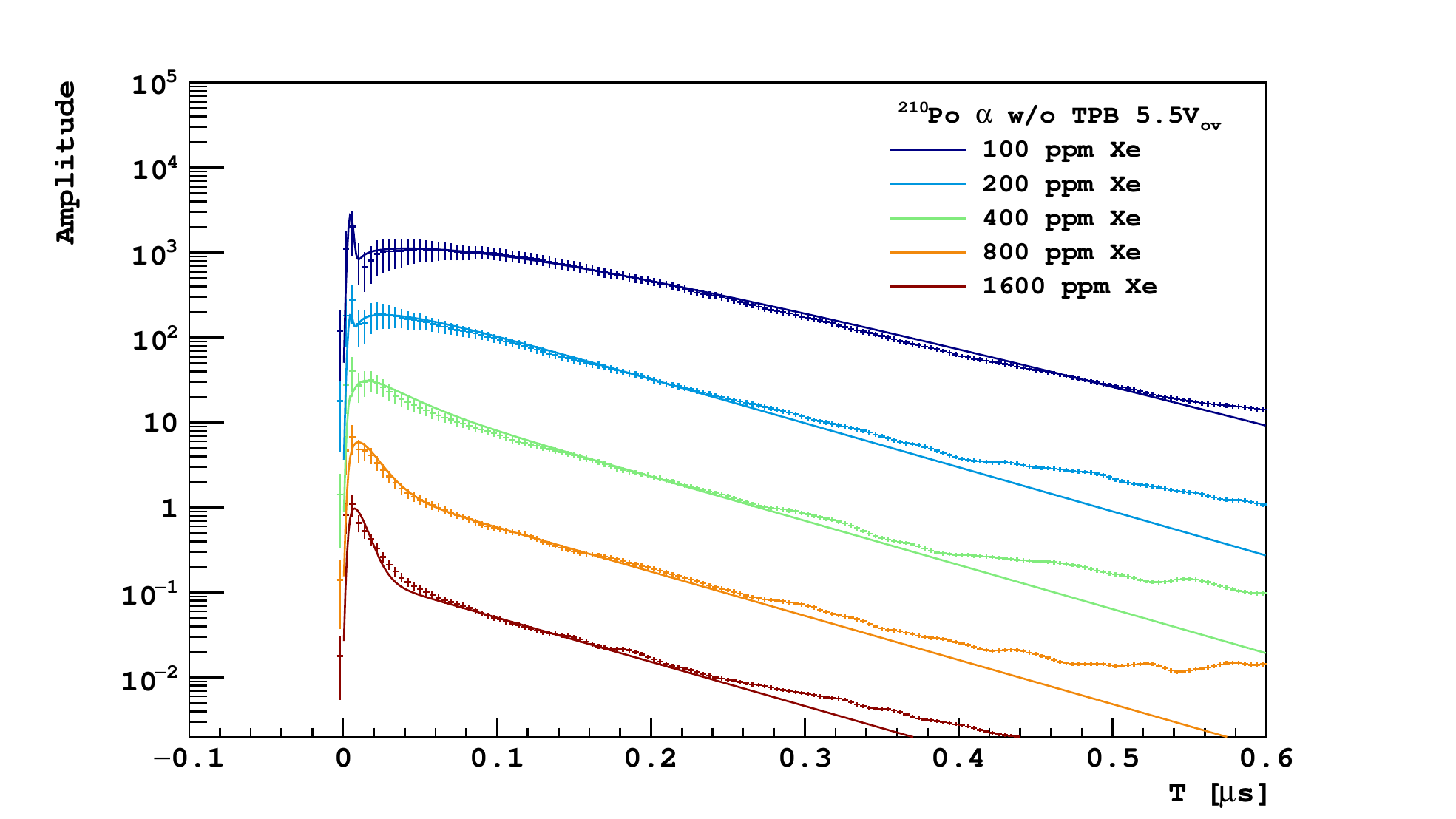}}
\subfigure[]{ \label{fig:ps:210PoTPB30p5v} 
\includegraphics[width=7.3cm, trim=0.8cm 0.2cm 2.1cm 0.5cm, clip=true]{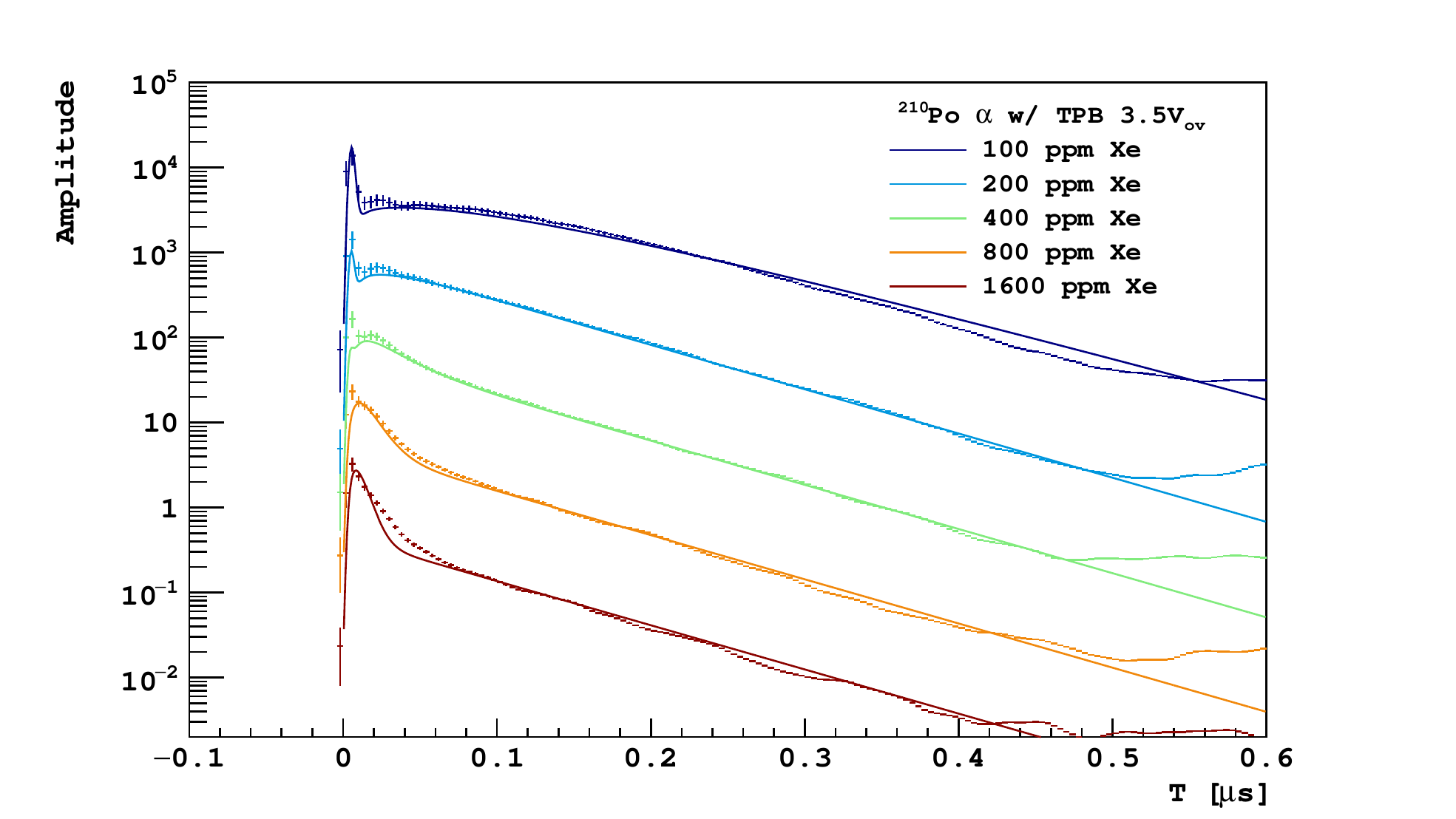}} 
\subfigure[]{ \label{fig:ps:210PoTPB32p5v} 
\includegraphics[width=7.3cm, trim=0.8cm 0.2cm 2.1cm 0.5cm, clip=true]{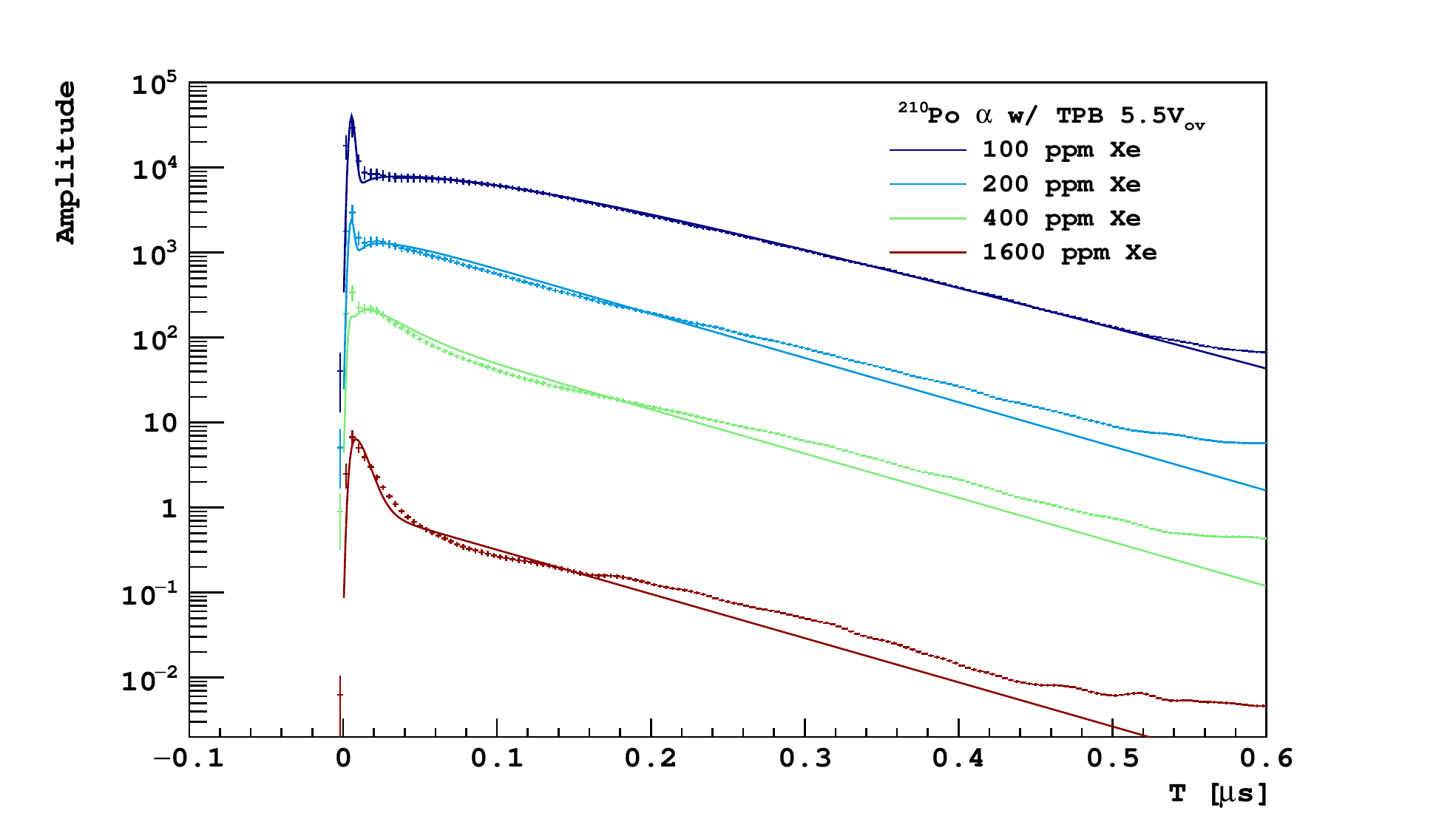}}
\subfigure[]{ \label{fig:ps:90Sr30p5v} 
\includegraphics[width=7.3cm, trim=0.8cm 0.2cm 2.1cm 0.5cm, clip=true]{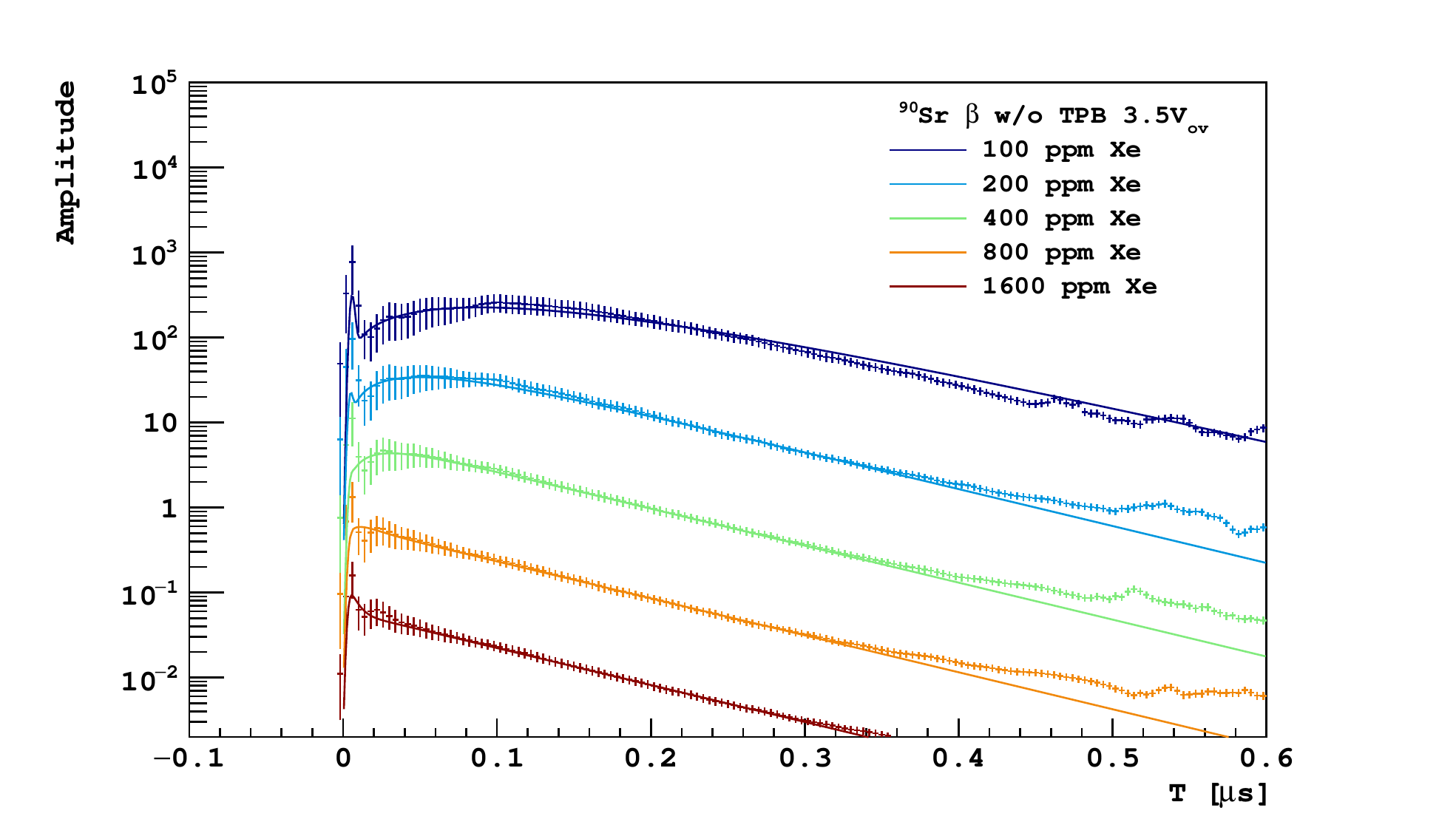}} 
\subfigure[]{ \label{fig:ps:90Sr32p5v} 
\includegraphics[width=7.3cm, trim=0.8cm 0.2cm 2.1cm 0.5cm, clip=true]{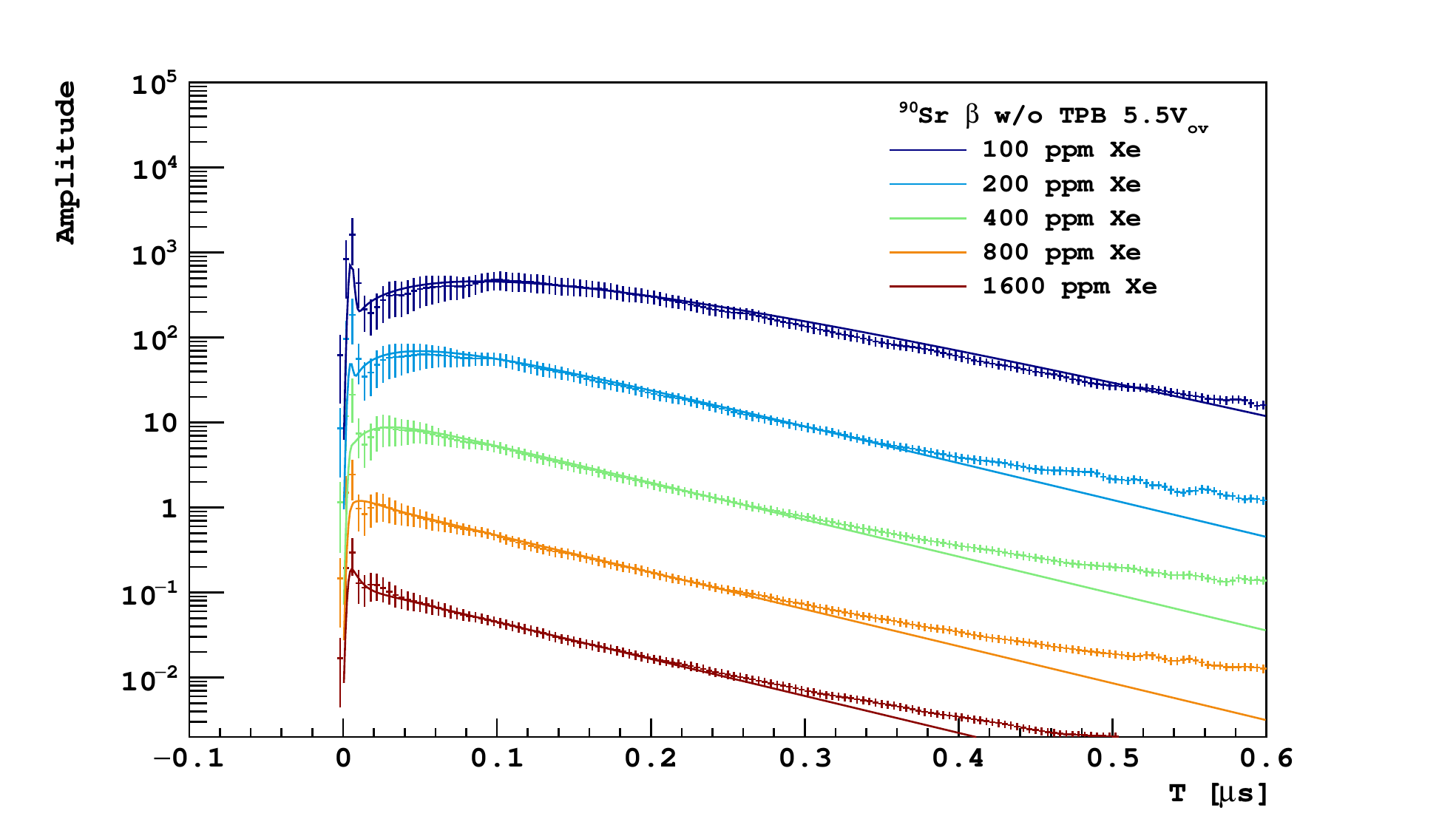}}
\caption{Model fit to the deconvoluted waveforms acquired under different test configurations. The histograms are deconvoluted average raw waveforms, and the solid line indicates the fit model. There is a factor of 10 offset between consecutive curves in y-axis for better visualization.} 
\label{fig:ps} 
\end{figure}

\begin{figure}
    \centering
    \includegraphics[width=\linewidth]{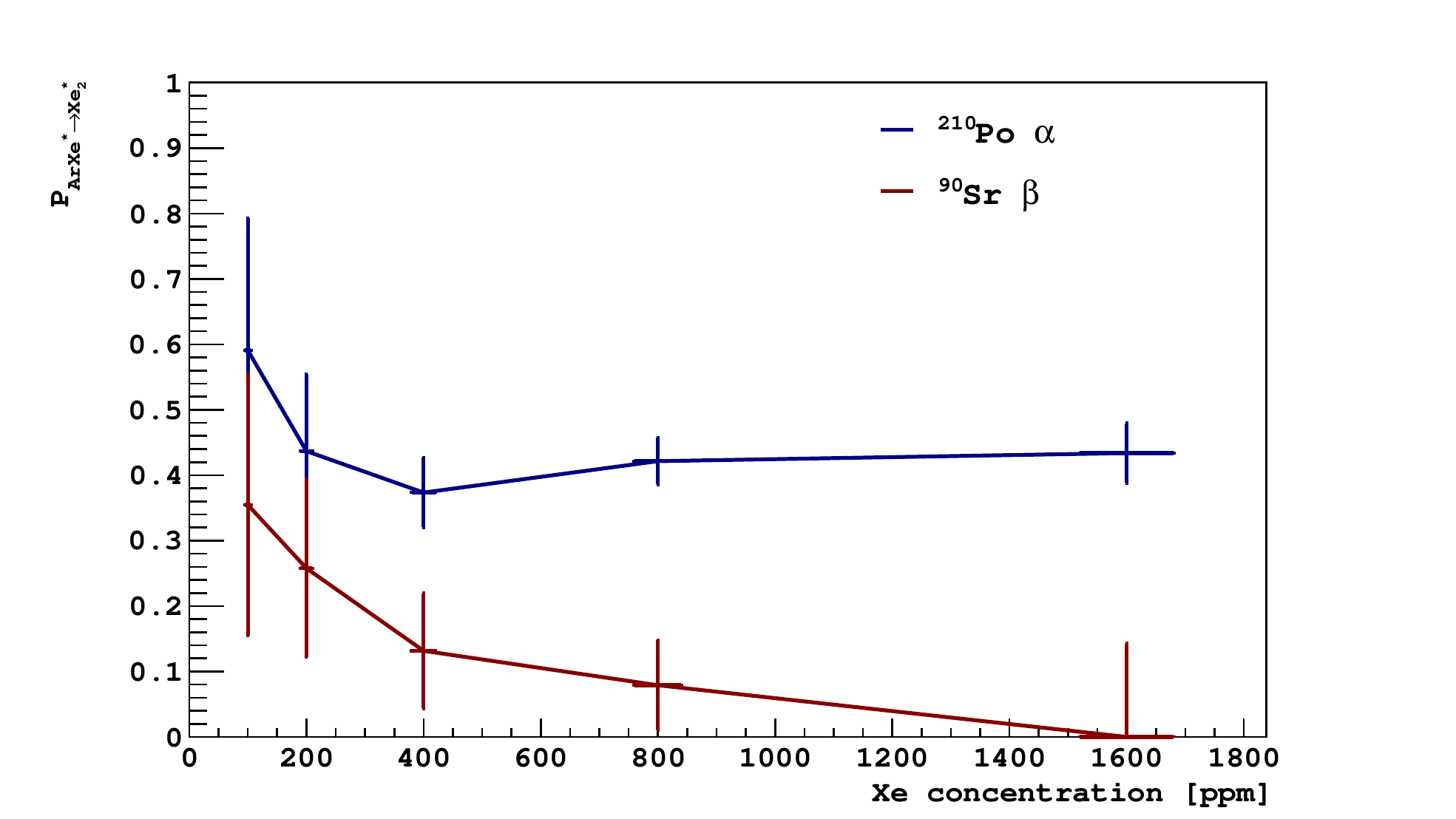}
    \caption{The prompt fraction $p_{\Xe}$ of $\Ar\Xe^{*}$ to $\Xe_{2}^{*}$.}
    \label{fig:PXe}
\end{figure}

The fitting results are shown in figure~\ref{fig:ps}, and the value of the parameters are listed in table~\ref{tab:pars}. 
The fitting result for the fraction of $\Ar\Xe^*\to\Xe_2^{*1}\Sigma$ with respect to the total $\Ar\Xe^*\to\Xe_2^*$ rate is shown in figure~\ref{fig:PXe}. 
In figure~\ref{fig:ps}, results of different configurations are plotted separately. 
From top to bottom, the plots are for $^{210}\mathrm{Po}~\alpha$ with TPB, $^{210}\mathrm{Po}~\alpha$ without TPB and $^{90}\mathrm{Sr}~\beta$ without TPB. 
In each subfigure, from blue to red, the xenon concentrations are \SI{100}{ppm}, \SI{200}{ppm}, \SI{400}{ppm}, \SI{800}{ppm} and \SI{1600}{ppm}, and the amplitude is shifted downwards by a factor of 10 sequentially. 
The histograms are the data with error bars estimated from the fluctuations in individual waveforms, and solid smooth curves denote the fitting model. 
Comparing to the pulse shape of $^{90}\mathrm{Sr}~\beta$ and $^{210}\mathrm{Po}~\alpha$, also shown in figure~\ref{fig:PXe}, the PSD power does not diminish at high xenon concentration, while the relatively short lifetime of the slow component makes the mixture more suitable in high event rate applications.
The slow component attributes to the decay time of $\Xe_{2}^{*3}\Sigma$, which is slower than the one in pure liquid xenon. 
A similar phenomenon has also been reported in~\cite{akimov2019}.
The extension of the decay time could be also effected by some other factors such as temperature~\cite{schwentner1985electronicInCondensedRG,kink1977solidxenon}, refractive index of the medium~\cite{morikawa1989mediumEffect,shibuya1983refractive} and the molecular density~\cite{leichner1976_2and3bodyRate,keto1974productionofXeArUV}. Further investigation is required to explain this observation.

\begin{table}
    \centering
    \begin{tabular}{c|c|c}
         \hline
         Parameter      & Bests Fit & Reference                          \\
         \hline 
         $k_{\Ar_{2}^{*}{}^{1}\Sigma,\Xe^{*}}$ [\SI{}{cm^3/s}]   & $3.9_{-1.6}^{+6.6}\times10^{-10}$  & \SI{3.3e-11}{} \cite{hitachi1993photon}\\
         $l_{\Xe^{*},\Ar\Xe^{*}}$ [\SI{}{\us^{-1}}]               & $1.2_{-0.5}^{+8.8}\times10^{6}$   & - \\
         $k_{\Ar\Xe^{*},\Xe_{2}^{*}}$ [\SI{}{cm^3/s}]            & $5.9_{-1.3}^{+1.7}\times10^{-12}$   & \SI{6e-12}{} \cite{cheshnovsky1973LKrXe}\\
         $\tau_{\Xe_2^*{}^3\Sigma} ~^{90}\mathrm{Sr} ~\beta$  [\SI{}{ns}]    & $100\pm\SI{6}{}$ & \multirow{2}{*}{22 \cite{hitachi1983IonDensityPSD} }\\
         $\tau_{\Xe_2^*{}^3\Sigma} ~^{210}\mathrm{Po} ~\alpha$ [\SI{}{ns}]   & $84\pm\SI{2}{}$ & \\
         $\mathrm{PDE}_{\SI{178}{nm}}/ \mathrm{PDE}_{\SI{420}{nm}} \SI{3.5}{V}$ & $0.088\pm0.003$ & \\
         $\mathrm{PDE}_{\SI{178}{nm}}/ \mathrm{PDE}_{\SI{420}{nm}}\SI{5.5}{V}$ & $0.076\pm0.003$ & \\
         \hline
    \end{tabular}
    \caption{Fit results for free parameters. }
    \label{tab:pars}
\end{table}

The reduced $\chi^2$ of the combined fit is \SI{2.1}{}. 
The poor $\chi^2$ might be the result of data processing, for instance, the deconvolution is not able to reproduce the true light time profile.
The model described in equation~\ref{eq:process} also does not reflect the complete processes taken place in this mixture. 
In the case of $^{90}\mathrm{Sr}~\beta$ source, we considered Cherenkov light is a possible contribution to the prompt component. 
But based on a toy Monte Carlo with \texttt{Geant4}, the average number of Cherenkov photon per $\SI{2.5}{MeV}~\beta$ in liquid argon is about $\SI{6e2}{}$, which is negligible comparing to the amount of light from scintillation, which is about $\SI{1.1e5}{}$ in total and $\SI{1.5e4}{}$ in the first $\SI{4}{ns}$. 

\begin{figure}
    \centering
    \includegraphics[width=\linewidth]{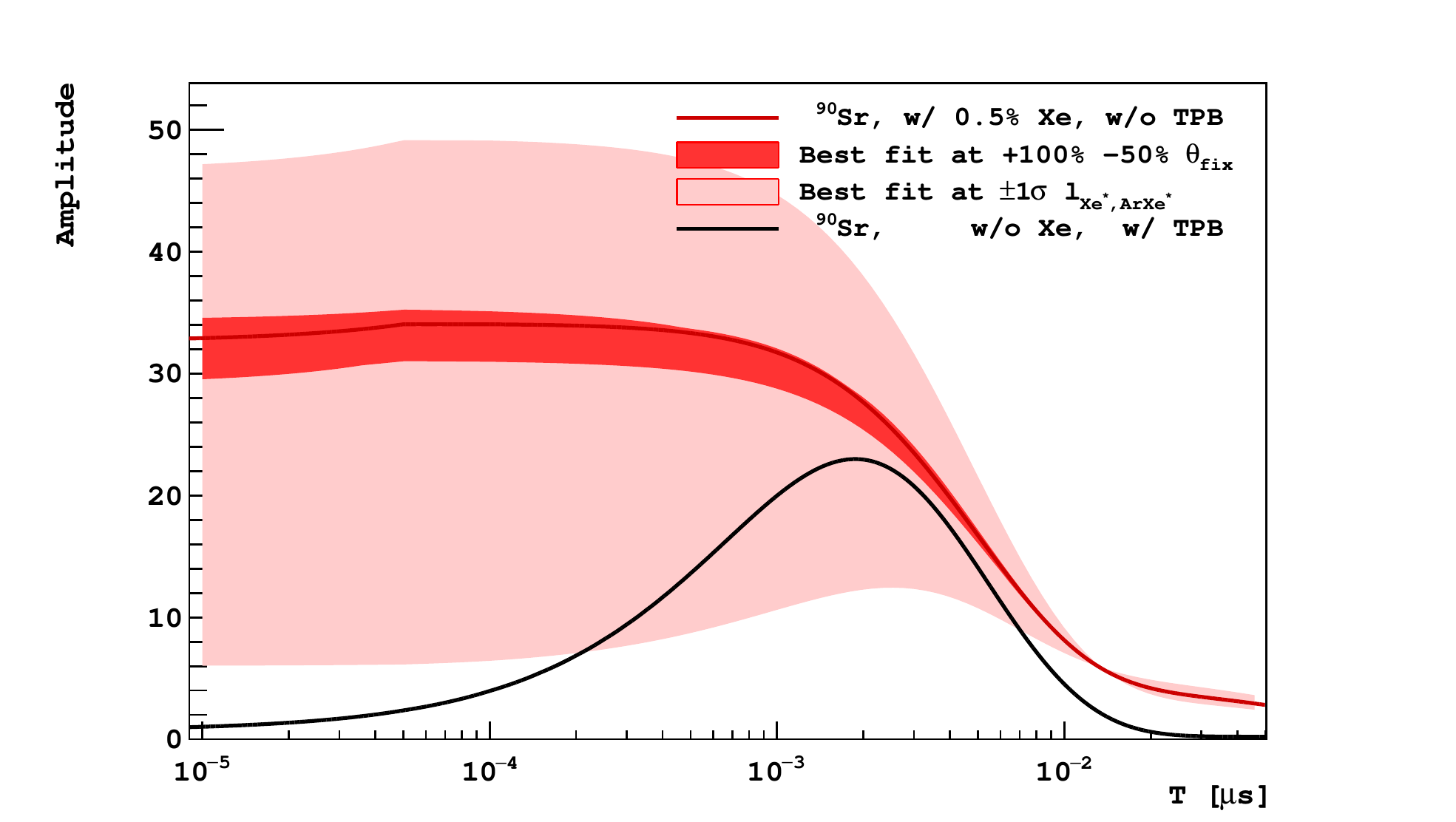}
    \caption{Simulated pulse from pure liquid argon (black) with TPB and liquid argon with \SI{0.5}{\percent} xenon (red). The PDE of SiPM to \SI{178}{nm} and \SI{420}{nm} wavelength light are assumed to be \SI{30}{\percent} and \SI{50}{\percent}, respectively. The dark red shade indicates the best fits with the first 5 parameters in table~\ref{tab:fixpars} varied between +\SI{100}{\percent} and -\SI{50}{\percent} individually. The light red shade indicates the best fits at $\pm1\sigma$ of $l_{\Xe^*,\Ar\Xe^*}$.}
    \label{fig:0p5percent}
\end{figure}

Nevertheless, this model effectively describes the trend of pulse-shape shifting with respect to the concentration of xenon, and the fit performs well with most of parameters taken from the literature.
Based on the fit results, for the fastest path marked as thick arrows in figure~\ref{fig:process}, when there is \SI{0.5}{\percent} of Xe in the mixture, the total time constant before photon emission is about \SI{10}{ps}, which is much faster than the case with TPB.
Figure~\ref{fig:0p5percent} shows the extrapolated pulse shape with SiPM of matched PDE to xenon scintillation light.
The PDE to \SI{178}{nm} wavelength light is assumed to be \SI{30}{\percent}, which is foreseeable with the current development of this technology.
In comparison, the PDE to \SI{420}{nm} wavelength light is assumed to be \SI{50}{\percent}.
In order to validate the robustness of the fit, the fit is repeated with parameters fixed at different values. 
The dark red shade represents the envelope of the best fit curves with parameters $\theta_\mathrm{fix}=\{l_{\Ar^*,\Ar^*_2}, k_{\Ar*,\Xe^*}, k_{\Ar^*_2{}^3\Sigma,\Xe^*}, k_{\Xe^*,\Xe^*_2{}^1\Sigma}, k_{\Xe^*,\Xe^*_2{}^3\Sigma}\}$ fixed to $+\SI{100}{\percent}$ and $-\SI{50}{\percent}$ of their values in table~\ref{tab:fixpars} individually. 
The light red shade represents the best fit with $l_{\Xe^*,\Ar\Xe^*}$ at $\pm1\sigma$.
The upper bound corresponds to the $-1\sigma$ fluctuation and the lower bound corresponds to the $+1\sigma$ fluctuation.
The fitting results agree with our expectations.
In figure~\ref{fig:process}, it is clear that the path of energy diverges at the node $\Xe^*{}^3\mathrm{P}_1$.
The thick arrow pointing to the xenon excimers are the fastest route, while the arrow pointing to the heteromolecular excimer $\Ar\Xe^*$ is competing with it by the rate constant $l_{\Xe^*,\Ar\Xe^*}$.
With larger $l_{\Xe^*,\Ar\Xe^*}$, more energy is captured and temporarily stored in $\Ar\Xe^*$, resulting in slower pulses.
The uncertainties on $k_{\Ar^*_2{}^1\Sigma,\Xe^*}$ and $k_{\Ar\Xe^*,\Xe^*_2}$ corresponds to relatively small variances on the extrapolated curve, as shown in table~\ref{tab:variance}.

\begin{table}
    \centering
    \begin{tabular}{c|c|c}
         \hline
         Parameter      & Pulse amplitude at $+1\sigma$ & Pulse amplitude at $-1\sigma$                        \\
         \hline 
         $k_{\Ar_{2}^{*}{}^{1}\Sigma,\Xe^{*}}$ & $-\SI{7}{\percent}$ & $\SI{4}{\percent}$\\
         $l_{\Xe^{*},\Ar\Xe^{*}}$ & $-\SI{80}{\percent}$ & $\SI{44}{\percent}$ \\
         $k_{\Ar\Xe^{*},\Xe_{2}^{*}}$ & $-\SI{12}{\percent}$   & $\SI{16}{\percent}$\\
         \hline
    \end{tabular}
    \caption{The variance of the predicated pulse amplitude at $T=\SI{10}{ps}$. }
    \label{tab:variance}
\end{table}

\section{Conclusion and outlook}

We have performed a detailed pulse shape study of the scintillation light produced by xenon-doped liquid argon with different concentrations.
The time profile of each energy transfer step during the scintillation process has been calculated by fitting the pulses.
The trend of the pulse shape shifting when adding xenon is observed, and it agrees to the previous studies~\cite{hitachi1983IonDensityPSD, hitachi1993photon, cheshnovsky1973LKrXe}.

This work is mainly limited by the PDE of the SiPM, especially the PDE of the VUV light when TPB is not present in the system.
Therefore, the timing resolution measurement with TPB is always better than the one without TPB.
From the fit, it is apparent that using xenon as the WLS, \SI{0.5}{\percent}  by mole fraction, can produce faster wavelength shifting than TPB.
While the evolution of SiPM technology is providing us promising PDE to VUV light, xenon-doped liquid argon can be eventually used for applications which require sub-nanosecond timing resolution.
For applications such as 3D TOF-PET, the timing is always determined by the first few photons emitted during the decay process. 
How to effectively capture these photons and study the limit of the timing resolution will be our next task.

\acknowledgments
We wish to thank Princeton’s Office of the Dean for Research for the support from the New Ideas for the Natural Sciences program.

\appendix
\section{Consideration of the free parameters}
\label{freepara}

$\mathrm{PDE}_{\SI{178}{nm}}/\mathrm{PDE}_{\SI{420}{nm}}$: the SiPM PDE to xenon scintillation light (\SI{178}{nm} as the wavelength) relative to the PDE to \SI{420}{nm} wavelength light. 

$\tau_{\Xe_2^*{}^3\Sigma}$:
The $\tau_{\Xe_2^*{}^3\Sigma}$ are set as two free parameters independently for $^{90}\mathrm{Sr}~\beta$ and $^{210}\mathrm{Po}~\alpha$ events.
In figure~\ref{fig:ps}, the scintillation light decay rates between \SI{0.2}{\us} and \SI{0.4}{\us} are very close to each other when the xenon concentration is more than \SI{100}{ppm}. 
This indicates that the slow component is independent of the xenon concentration in these cases.
According to the process shown in figure~\ref{fig:process}, the only slow process that is independent of the xenon concentration is the decay of $\Xe_{2}^{*}{}^3\Sigma$. 
From the fit, the decay time is around \SI{100}{ns}, which is slower than the rate in pure liquid xenon, and close to the value in gaseous xenon~\cite{leichner1976_2and3bodyRate}.
This result agrees with previous measurements~\cite{kubota1993suppression}.

$p_{\Xe}$: the fraction of singlet state in xenon excimers in the process of $\Ar\Xe^*\to\Xe_2^*$. They are set as independent free parameters for different sources and xenon concentrations, but keep the same value when changing only the TPB status or the SiPM bias voltages.

$p_{\Ar}$: the fraction of singlet state in argon excimers. As a xenon atom immediately interacts with $\Ar^{*}$ during the relaxation process before the formation of $\Ar_{2}^{*}$, this value could be different from pure liquid argon, but in order to avoid over-fitting, $p_{\Ar}$ is fixed to the pure liquid argon value. 

The relative pulse amplitude of $^{210}\mathrm{Po} ~\alpha$ and $^{90}\mathrm{Sr}~\beta$ events is also set as a free parameter. Since it is related to the event selection cut and not relevant to the pulse shape, the result is not reported.

\end{document}